\documentclass[aps, prd, onecolumn, tightenlines,notitlepage, superscriptaddress, nofootinbib,preprintnumbers, floatfix,showkeys]{revtex4-1}

\usepackage{amstext}
\usepackage{amssymb}
\usepackage{amsmath}
\usepackage{graphicx}
\usepackage{url}
\usepackage{color}
\usepackage[utf8]{inputenc}
\pdfoutput=1
\usepackage{textcomp}
\usepackage{url}

\usepackage{epsfig,amsfonts,mathrsfs,amsmath,amssymb,graphicx,color,slashed,multirow}
\usepackage{amsmath,latexsym,amssymb,graphicx,slashed,color,enumerate,url,cancel,gensymb}
\usepackage{textcomp}

\usepackage[x11names]{xcolor}
\usepackage{hyperref} 
\hypersetup{
    colorlinks=true,
    linkcolor=blue,
    urlcolor=blue,
    linktoc=blue,
    citecolor=red
           }
\usepackage[normalem]{ulem}

\definecolor{nicered}{rgb}{0.7,0.1,0.1}
\definecolor{nicegreen}{rgb}{0.1,0.5,0.1}

\usepackage{comment}

\begin{document}

\title{
 Probing nuclear properties and neutrino physics with current and future CE$\nu$NS experiments
}

\author{R. R. Rossi}  \email{rrossi@ifi.unicamp.br}
\affiliation{Instituto de F\'isica Gleb Wataghin - Universidade Estadual de Campinas, {13083-859}, Campinas SP, Brazil}

\author{G. Sanchez Garcia}\email{gsanchez@ific.uv.es}
\affiliation{Instituto de Física Corpuscular, CSIC-Universitat de València, and Departament de Física Teòrica, Universitat de València, C/Catedrático José Beltrán 2, Paterna 46980, Spain}

\author{M. Tórtola}\email{mariam@ific.uv.es}
\affiliation{Instituto de Física Corpuscular, CSIC-Universitat de València, and Departament de Física Teòrica, Universitat de València, C/Catedrático José Beltrán 2, Paterna 46980, Spain}

\begin{abstract}

The recent observation of Coherent Elastic Neutrino Nucleus Scattering (CE$\nu$NS) with neutrinos  from pion decay at rest ($\pi$-DAR) sources by the COHERENT Collaboration has raised  interest in this process in the search for new physics. 
Unfortunately, current uncertainties in the determination of nuclear parameters relevant to those processes can hide new physics effects.
This is not the case for processes involving lower-energy neutrino sources such as nuclear reactors. Note, however, that a CE$\nu$NS measurement with reactor neutrinos depends largely on the determination of the quenching factor, making its observation more challenging.
In the upcoming years, once this signal is  confirmed, a combined analysis of $\pi$-DAR and reactor CE$\nu$NS  experiments will be very useful to probe particle and nuclear physics, with a reduced dependence on the nuclear uncertainties. 
In this work, we explore this idea by simultaneously testing the  sensitivity  of current and future CE$\nu$NS experiments to neutrino non-standard  interactions (NSI) and the neutron
root mean square (rms) radius, considering different neutrino sources as well as several detection materials. 
We show how the interplay between future reactor and accelerator CE$\nu$NS experiments can help to get robust constraints on the neutron rms, and to break degeneracies between the NSI parameters.
Our forecast could be used as a guide to optimize the experimental sensitivity to the parameters under study.\\

\end{abstract}

\maketitle

\section{Introduction}

The process of Coherent Elastic Neutrino-Nucleus Scattering (CE$\nu$NS) was theoretically proposed more than 40 years ago~\cite{PhysRevD.9.1389}. In this process, a relatively low-energy neutrino interacts with a nucleus as a whole and, as a result of the interaction, the nucleus acquires a kinetic recoil energy that can be measured. 
Given its low energy signal, CE$\nu$NS  was not observed until 2017, when the COHERENT collaboration reported its first experimental measurement by using 
 neutrinos from the Spallation Neutron Source (SNS) and a cesium iodide (CsI) detector~\cite{COHERENT:2017ipa} at Oak Ridge National Laboratory (ORNL).
A second measurement was performed by the same collaboration in 2020~\cite{COHERENT:2020iec}, this time by using a liquid argon (LAr) detector, and a more recent dataset from the CsI detector was also released in 2021~\cite{COHERENT:2021xmm}. 
Since then, CE$\nu$NS has been widely used to test Standard Model (SM) parameters \cite{AtzoriCorona:2023ktl, Cadeddu:2021ijh}, to study nuclear physics parameters through the neutron
root mean square (rms) radius~\cite{Cadeddu:2017etk, Papoulias:2019lfi}, as well as to constrain new physics scenarios. These include neutrino non-standard interactions (NSI)~\cite{Ohlsson:2012kf, Denton:2020hop, Miranda:2020tif, Giunti:2019xpr, Miranda:2019skf}, neutrino  electromagnetic properties~\cite{Miranda:2019wdy, Kosmas:2015sqa, Cadeddu:2018dux, Parada:2019gvy}, neutrino generalized interactions (NGI) \cite{Lindner:2016wff, AristizabalSierra:2018eqm, Flores:2021kzl}, CP-violating effects \cite{AristizabalSierra:2019ufd}, light mediators \cite{Abdullah:2018ykz, Cadeddu:2020nbr, Flores:2020lji}, sterile neutrinos \cite{Kosmas:2017zbh, Blanco:2019vyp}, and dark fermion production~\cite{Candela:2023rvt}, among many other scenarios. 
A recent combined analysis of COHERENT CsI and LAr data can be found in Ref.~\cite{DeRomeri:2022twg}.

The coherent character of CE$\nu$NS comes from the fact that, for energies up to some tens of MeV, the contribution to the cross section from the neutrons inside the nucleus adds up coherently, giving as a result a characteristic quadratic dependence on the number of neutrons of the target material. Because of the incoming neutrino energy range needed for coherency, neutrinos from pion decay at rest ($\pi$-DAR) sources are suitable to study CE$\nu$NS. 
This is the case of the COHERENT collaboration, which uses neutrinos produced at the SNS \cite{COHERENT:2021yvp}. In addition, many
other collaborations aim to measure CE$\nu$NS from different $\pi$-DAR sources in the following years, such as the CCM experiment~\cite{CCM:2021leg}, as well as the experimental proposal at the future European Spallation Source (ESS)~\cite{Baxter:2019mcx}.

After the two successful measurements of CE$\nu$NS by COHERENT, there has been a wide interest from the community in measuring CE$\nu$NS by using different target materials and neutrino sources. For instance, using different materials can help to unambiguously corroborate the quadratic dependence of the cross section on the number of neutrons of the target material. In fact, the complete COHERENT program includes detectors from different technologies such as Ge and NaI~\cite{Akimov:2022oyb}. 
As an alternative, the feasibility of using different isotopes of the same material to test the CE$\nu$NS cross section has also been proposed in~\cite{Galindo-Uribarri:2020huw}. 
This approach can also be useful in the searches for new physics with CE$\nu$NS.
 It has been shown, for example,  that degeneracies in the determination of neutrino non-standard interactions (NSI) with matter can be lifted when combining CE$\nu$NS results from different detectors~\cite{chatterjee2023constraining}. 
As for practical purposes, the applicability of CE$\nu$NS for reactor monitoring has also been studied~\cite{vonRaesfeld:2021gxl}. 

Likewise,  the interest in measuring CE$\nu$NS from different neutrino sources has been increasing throughout the years. 
Indeed, very low thresholds in Dark Matter (DM) detectors will soon become sensitive to CE$\nu$NS from solar neutrinos as part of their backgrounds.
However, the most promising channel for CE$\nu$NS detection through new neutrino sources is that of reactor neutrinos. These neutrinos are in a lower energy regime than those coming from $\pi$-DAR sources, and hence, their detection is significantly important. In the case of $\pi$-DAR neutrinos, the cross section is sensitive to the nuclear structure of the target material through the form factor. In contrast, for reactor neutrino energies, nuclear distribution effects do not play a significant role and so they are negligible (see Sec.~\ref{sec:source}). This feature can be exploited to obtain complementary measurements that can help to study new physics effects without the uncertainties that come from nuclear effects.

Regarding the current status of CE$\nu$NS searches using reactor neutrinos, suggestive evidence of a positive signal has been reported in \cite{Colaresi:2022obx}. Note, however, that its interpretation is highly dependent on quenching factor measurements. Beyond this result,  many experimental efforts are currently under development with the aim of measuring CE$\nu$NS from reactor neutrinos, as is the case of experiments like CONUS~\cite{CONUS:2020skt}, $\nu$GeN~\cite{nGeN:2022uje}, CONNIE~\cite{CONNIE:2021ggh}, and Red-100~\cite{Akimov:2022xvr}, among many others.
Then,  a detailed forecast of the complementarity between current and future experiments of different neutrino sources is needed. In this work, we explore how the combination of CE$\nu$NS experiments that use neutrinos produced from different sources could help us to constrain both NSI and nuclear parameters. This kind of analysis was first explored in \cite{canas2020interplay} for a single flavor-conserving NSI parameter. Here we extend the analysis to both non-universal and flavor-changing NSI. In the case of $\pi-$DAR sources, we first make use of the latest data from the COHERENT CsI detector. In addition, we explore the expected sensitivity to the same parameters of two future $\pi-$DAR source experiments: a Ge detector located at the SNS, and a proposed Xe detector at the future ESS. 
As for reactor neutrinos, we explore their near-future expected sensitivity to NSI by considering the detector characteristics and background models of a CONUS-like~\cite{CONUS:2020skt} and a $\nu$GeN-like experiment~\cite{nGeN:2022uje}, each under a different quenching factor (QF) assumption, since the observation of CE$\nu$NS for reactor neutrinos largely depends on this quantity. 
 With this approach, our estimation of the sensitivity of future CE$\nu$NS experiments will be as realistic as possible.

 The remainder of this paper is organized as follows. In Sec. \ref{sec:cevns:theo} we discuss the theoretical features of CE$\nu$NS, the role of the rms radius, and the general framework of NSI. 
 In Sec.~\ref{sec:source} we describe in detail the different neutrino sources considered in this work, as well as the characteristics of the analyzed neutrino detectors. 
 In Sec.~\ref{sec:analysis}  we discuss the analysis procedure followed to obtain the sensitivities on the nuclear physics and new physics scenarios investigated, with the corresponding results presented in Sec.~\ref{sec:results}. 
 Finally,  we draw our conclusions in Sec.~\ref{sec:conclusions}. 

\section{Theoretical framework}\label{sec:cevns:theo}

\subsection{Coherent elastic neutrino-nucleus scattering}\label{sec:cevns}

Within the SM, the CE$\nu$NS cross section is given by \cite{PhysRevD.9.1389}
\begin{equation}
\label{eq:cross}
\left (  \frac{\mathrm{d}\sigma}{\mathrm{d}T} \right )_{\textrm{SM}} = \frac{G_F^2M}{\pi}\left(1-\frac{MT}{2E_{\nu}^2}\right)\,\left(Q_{W}^V\right)^2,
\end{equation}
where $G_F$ is the Fermi constant, $M$ is the mass of the target material, $T$ is the nuclear recoil energy, $E_\nu$ is the energy of the incoming neutrino, and $Q_W^V$ is the weak charge, given by
\begin{equation}\label{eq:qweak}
    \left(Q_{W}^V\right)^2  = \left(ZF_Z(q^2)\,g_V^p + NF_N(q^2)\,g_V^n \right)^2,
\end{equation}
with $g_V^p = 1/2 - 2\sin^2{\theta_W}$ and $g_V^n = -1/2$ the coupling constants defined in the SM, and $\theta_W$ the weak mixing angle. Notice that within the SM, the CE$\nu$NS cross section is flavor independent, with small corrections that have been studied in \cite{Tomalak:2020zfh} but that are not relevant for current experimental sensitivities. The functions $F_Z(q^2)$ and $F_N(q^2)$ in Eq.~(\ref{eq:qweak}) are called the proton and neutron form factors, respectively, and they describe the distribution of the corresponding protons and neutrons within the nucleus as a function of the momentum transfer, $q$. We can find different parametrizations to these form factors in the literature, including the symmetrized Fermi \cite{Piekarewicz:2016vbn}, the Helm \cite{PhysRev.104.1466}, and the Klein-Nystrand \cite{PhysRevC.60.014903} parametrizations. Our results are independent of the parametrization, and we use the Klein-Nystrand one, which is given by
\begin{equation}\label{eq:KNFF}
   F_{X}(|\vec{q}|^2)=3\frac{j_1(|\vec{q}|R_X)}{|\vec{q}|R_X} \left(\frac{1}{1+|\vec{q}|^2a_k^2} \right)\ ,
\end{equation}
with $X = Z, N$ standing for protons and neutrons, respectively. In this equation, $j_1$ is the spherical Bessel function of order 1, $a_k = 0.7$ fm, and each $R_X$ satisfies \cite{Sierra:2023pnf}

\begin{equation}
    R_Z^2 = \frac{5}{3}\left ( R_p^2-6a_k^2 \right ),
\end{equation}
and
\begin{equation}
    R_N^2 = \frac{5}{3}\left ( R_n^2-6a_k^2 \right ),
    \label{eq:rms:n}
\end{equation}
with $R_p$ and $R_n$ the proton and neutron rms radius, respectively.  Thanks to the electromagnetic coupling of protons, there are many experimental measurements of the parameter $R_p$ for different materials \cite{Angeli:2013epw}. In contrast, the neutron rms radius is only known for a few materials, and the process of CE$\nu$NS can be used for its determination \cite{Cadeddu:2017etk}.

\subsection{Non-Standard Interactions}

The formalism of NSI accounts for possible new physics at low energies in the lepton sector of the SM, and may affect neutrino interactions in their production, propagation, or detection. Here we focus on the case of NSI at the detection point, which was first studied in \cite{PhysRevD.17.2369}. In general, NSI can be either of charged or neutral current character. For the case of  neutral current NSI, our topic of interest, the Lagrangian that needs to be added to the SM, parametrized in terms of the Fermi constant, is given by
\begin{equation}
{\cal{L}}^\mathrm{NSI}_\mathrm{NC}=-2\sqrt{2}G_F \sum\limits_{\alpha,\beta} \varepsilon_{\alpha\beta}^{f C}(\bar{\nu}_\alpha\gamma^\mu P_L\nu_{\beta})(\bar{f}\gamma_\mu P_C f) \, ,
\end{equation}
where greek characters run over the three neutrino flavors, $C$ = $L$, $R$ stands for the chirality, and $f$ represents any of the SM charged fermions. The NSI constants $\varepsilon_{\alpha\beta}^{f C}$ parametrize the new physics strength of interaction weighted by $G_F$. Then, we expect these interactions to be below $\cal{O}$({1}) so that they are subdominant with respect to the SM weak force. Since we are interested in the interaction of neutrinos with a nucleus, we will focus on the case where the fermions are the $up$ and $down$ quarks within the nucleus. Then, once  NSI are introduced for an incoming neutrino of flavor $\alpha$, the weak charge defined in Eq.~\eqref{eq:qweak} is modified to \cite{Barranco:2005yy}
\begin{eqnarray}
\left(Q_{W,\alpha}^{V,\textrm{NSI}}\right)^2 & = \left[Z\left(g_V^p + 2\varepsilon_{\alpha\alpha}^{uV}+\varepsilon_{\alpha\alpha}^{dV}\right) + N\left(g_V^n + \varepsilon_{\alpha\alpha}^{uV}+2\varepsilon_{\alpha\alpha}^{dV}\right)\right]^2 \nonumber \\
 & + \sum\limits_{\beta\neq \alpha} \left| Z\left(2 \varepsilon_{\alpha\beta}^{uV} + \varepsilon_{\alpha\beta}^{dV} \right)  + N\left(\varepsilon_{\alpha\beta}^{uV} + 2\varepsilon_{\alpha\beta}^{dV} \right) \right|^2
 \label{eq:weak_charge_NSIs}\, ,
\end{eqnarray}
where we have defined the vector coupling constants as \footnote{Axial contributions in CE$\nu$NS interactions are expected to be of order $1/A$ and, hence, they are negligible for the nuclei of interest.}
\begin{equation}
\varepsilon_{\alpha\beta}^{qV} = \varepsilon_{\alpha\beta}^{qR} + \varepsilon_{\alpha\beta}^{qL}\, .
\end{equation}
In contrast to the SM, where the weak force is universal and flavor conserving,  NSI can be either of a non-universal ($\alpha = \beta$) or a flavor changing ($\alpha \neq \beta$) character. For a  general review on neutrino NSI, we refer the reader to Ref.~\cite{Farzan:2017xzy}. 

\section{Neutrino sources for CE\lowercase{v}NS}\label{sec:source}

In this section, we briefly describe two neutrino sources that can be used for the study of CE$\nu$NS: stopped pion neutrinos and reactor neutrinos.  Although we focus on these sources, we would like to remind the reader that very low threshold DM experiments will be sensitive to the CE$\nu$NS contribution from solar neutrinos.
Besides discussing the relevant neutrino sources, in this work, we also discuss several potential detectors that could be used in future CE$\nu$NS searches. Our working scenarios can be regarded as futuristic or optimized versions of current detectors, such as COHERENT, CONUS, or $\nu$Gen.

\subsection{$\pi$-DAR sources}

Neutrinos from these sources are produced through the collision of high-energy proton beams with a large-density target material. Spallation sources represent a particular kind of $\pi$-DAR sources on which proton beams are delivered in pulses with a well-determined frequency. This feature plays an important role in the determination and characterization of backgrounds for CE$\nu$NS experiments. Here charged pions are produced as a byproduct of the proton collision. Then, pions are thoroughly stopped and they decay at rest producing muon neutrinos. Since this is a two-body decay, these neutrinos, often referred to as \textit{prompt} neutrinos, have a fixed energy and the corresponding flux can be described with a delta function
\begin{equation} 
\frac{d N_{\nu_\mu}}{d E_\nu}(E_\nu)  = \eta \, \delta\left(E_\nu-\frac{m_{\pi}^{2}-m_{\mu}^{2}}{2 m_{\pi}}\right)\,.
\label{eq:prompt:flux}
\end{equation}
Here $m_\pi$ is the mass of the pion, $m_\mu$ is the mass of the muon, and $\eta = rN_{\textrm{POT}}/4\pi L^2$  is a normalization factor that depends on the distance from the source to the detector, $L$, the number of protons on target, $N_{\textrm{POT}}$, and the number of neutrinos per flavor released during the interaction, $r$. As pions decay, there is also a production of muons, which eventually also decay at rest producing electron neutrinos and muon antineutrinos. Since this last decay is a three-body process, the energy of the released neutrinos, known as \textit{delayed} neutrinos, is no longer fixed, and it is described by the following distributions 
\begin{equation}
\frac{d N_{\bar{\nu}_\mu}}{d E_\nu}(E_\nu) = \eta \frac{64 E^{2}_\nu}{m_{\mu}^{3}}\left(\frac{3}{4}-\frac{E_\nu}{m_{\mu}}\right),
\label{eq:delayed:1}
\end{equation}
\begin{equation}
\frac{d N_{\nu_e}}{d E_\nu}(E_\nu)  = \eta \frac{192 E^{2}_\nu}{m_{\mu}^{3}}\left(\frac{1}{2}-\frac{E_\nu}{m_{\mu}}\right)  \, ,
\label{eq:delayed:2}
\end{equation}
where $\eta$ is the same normalization factor defined above. 
Then, the total flux at $\pi$-DAR sources will be the sum of the three neutrino contributions in Eqs.~(\ref{eq:prompt:flux}), (\ref{eq:delayed:1}) and (\ref{eq:delayed:2}). 
Once the cross section is known, the number of events for each neutrino flux component at a CE$\nu$NS experiment can be computed as 
\begin{align}\label{eq:Nevents_CEvNS_alpha}
N_{i,\alpha}^{\mathrm{CE}\nu\mathrm{NS}}
= 
\mathcal{N}
\int_{T_{i}}^{{T}_{i+1}}
\hspace{-0.3cm}
d T\,
\epsilon(T)
\int_{0}^{T^{\prime\text{max}}}
\hspace{-0.3cm}
dT'
\,
G(T,T')  \int_{E_\nu^{\text{min}}(T')}^{E_\nu^{\text{max}}}
\hspace{-0.3cm}
d E_\nu
\frac{d N_{\nu_\alpha}}{d E_\nu}(E_\nu)
\frac{d\sigma}{dT'}(E_\nu, T'),
\end{align}
where $i$ corresponds to the  $i$th energy bin and $\mathcal{N}$ is the number of targets inside the material. For a $\pi$-DAR experiment, $\mathcal{N} = N_Am_{\textrm{det}}/M_\textrm{mol}$, with $N_A$ the Avogadro's number, $m_{\textrm{det}}$ the mass of the detector, and $M_{\textrm{mol}}$ the molar mass of the material. The limits on the integral over $E_\nu$ are such that $E_\nu^{\textrm{min}} = \sqrt{MT'/2}$,  while $E_\nu^{\textrm{max}}$ corresponds to the maximum energy for the incoming neutrinos, which for $\pi$-DAR sources is approximately $52.8$ MeV. 
 The resolution of the detector is  accounted for by using the smearing function $G(T,T')$, which depends on the true recoil energy of the nucleus, $T'$, and the reconstructed energy, $T$. 
 In addition, Eq. \eqref{eq:Nevents_CEvNS_alpha} includes the efficiency of the detector $\epsilon(T)$.  The total number of events for an energy bin $i$, is given by the sum over the three neutrino flux components

 \begin{align}\label{eq:Nevents_CEvNS}
N_{i}^{\mathrm{CE}\nu\mathrm{NS}}
= \sum_{\alpha}N_{i,\alpha}^{\textrm{CE$\nu$NS}}
\end{align}

  In some cases, timing distributions for the measured CE$\nu$NS events are provided by the experiment. Accounting for this information when necessary, the number of events associated to an energy bin $i$ and a timing bin $j$, can be calculated as
\begin{align}\label{eq:time_CEvNS}
N_{ij}^{\mathrm{CE}\nu\mathrm{NS}}
= \sum_{\alpha}N_{i,\alpha}^{\textrm{CE$\nu$NS}}\int_{t_j}^{t_{j+1}}f_{\alpha}(t)\epsilon_{t}(t)dt,
\end{align}
where $f_{\alpha}(t)$ is the flavor-dependent timing distribution, and $\epsilon_t(t)$ is the timing efficiency. 
In this work, we consider the latest data set of the COHERENT CsI detector, as well as two future data samples involving $\pi-$DAR sources.
 For the latter, we simulate future results to be obtained by an undergoing Ge detector from the COHERENT collaboration at the SNS, and we analyze the sensitivity of a proposed Xe-based detector
at the still-under-construction ESS.

\subsubsection{COHERENT CsI detector at the SNS}

\begin{figure}[t]
    \centering
    \includegraphics[scale=0.5]{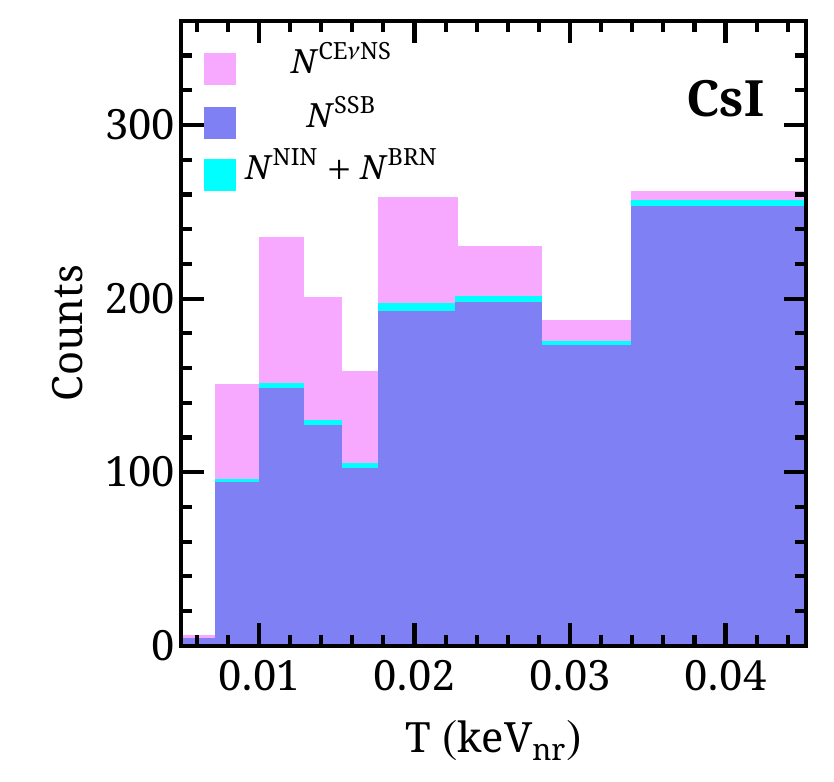}
    \includegraphics[scale=0.5]{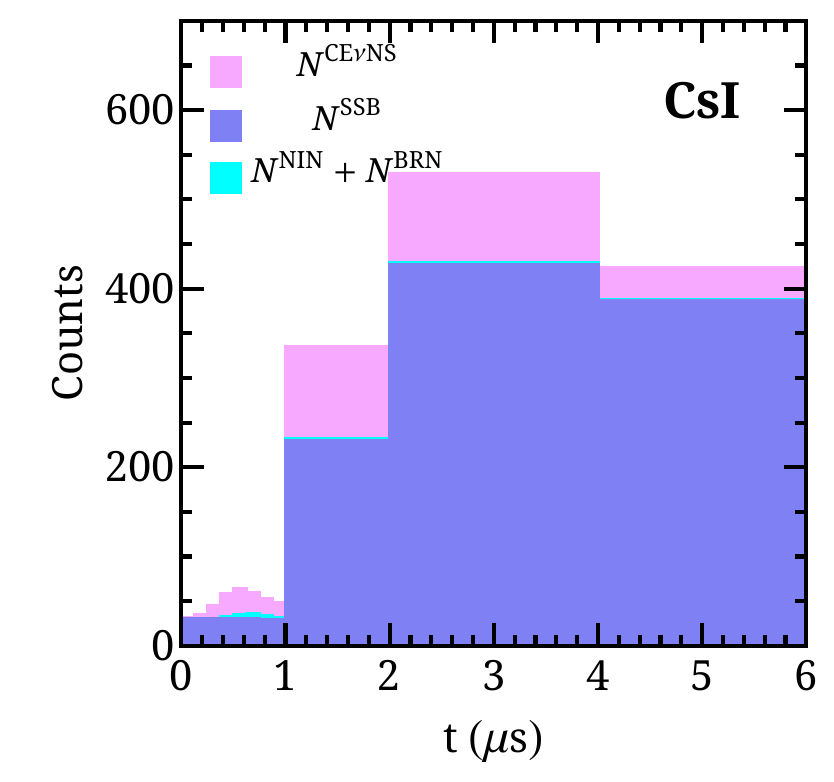}
    \caption{Predicted number of CE$\nu$NS events by the SM for the CsI detector used by the COHERENT collaboration. The left panel shows the results in the recoil energy, $T$, parameter space, while right panel shows the distribution in time. Besides CE$\nu$NS data (pink), the associated backgrounds are also indicated in the figure: SSB in purple, and NIN + BRN, in cyan.} 
    \label{fig:events:CsI}
\end{figure}

In 2021, the COHERENT collaboration reported an updated data set of the CsI detector used to measure CE$\nu$NS~\cite{COHERENT:2021xmm}. 
The CsI detector was located at 19.3 m from the neutrino source and, for the running time of the experiment, the collaboration reported a value of  $N_{POT} = 3.198\times10^{23}$, with a number of neutrinos per flavor of $r = 0.0848$ \cite{COHERENT:2021xmm}. 
These parameters are summarized in Tab. \ref{tab:snschar}, and allow us to compute the total neutrino flux expected at the experiment. 
The smearing function
$G(T, T')$, as well as the efficiency shape for the detector, are
provided by the collaboration in the supplemental materials of Ref.~\cite{COHERENT:2021xmm} as a function of the number of photoelectrons (PE) emitted by the scintillating material. 
These PEs result from the nuclear recoil energy and satisfy
\begin{equation}
    \textrm{PE} = \textrm{LY}\times T_{ee}\,,
\end{equation}
where LY is the light yield (13.35 PE/keV$_{ee}$ \cite{COHERENT:2017ipa}) and $T_{ee}$ is the electron-equivalent recoil energy. This quantity is related to the nuclear recoil energy, $T$, through 
\begin{equation}
    T_{ee} = \textrm{QF}\times T\,,
\end{equation}
where QF is the \textit{quenching factor}, defined as the fraction of kinetic recoil energy detected for the recoil of a heavy particle when compared to an incident electron of the same energy. For CsI, the electron equivalent energy, $T_{ee}$, is a function of the nuclear recoil energy $T$, with a polynomial of order four dependence, which is provided in the supplemental material in Ref.~\cite{COHERENT:2021xmm}, and from which one can extract the QF at a given energy. 
Finally, the COHERENT collaboration has also provided information on the timing distribution of CE$\nu$NS events. Then, to compute the expected number of events due to CE$\nu$NS we use Eq.~(\ref{eq:time_CEvNS}), with timing shape taken from \cite{Picciau:2022xzi} and the corresponding efficiency from \cite{COHERENT:2021xmm}. In Fig. \ref{fig:events:CsI}, we show a projection of the obtained expected number of events in the nuclear energy recoil parameter space (left panel) and in the timing space (right panel). In the same figure, we also show the backgrounds as provided by the COHERENT collaboration \cite{COHERENT:2021xmm}. 
For the SNS, the reported dominant contribution to them comes from Steady State Backgrounds (SSB), which are those coming from natural sources such as cosmic rays. In addition, there is a minor contribution from Beam Related Neutrons (BRN),  produced as the result of the original proton collisions, and Neutrino Induced Neutrons (NIN),  produced by charged and neutral current interactions of neutrinos with the lead shield surrounding the detector. The different bin widths in energy recoil in Fig. \ref{fig:events:CsI} are taken so that they match the PE width shown by the collaboration in Ref. \cite{COHERENT:2021xmm}.~\footnote{For more details of the binning distribution and background treatments, we refer the reader to Ref. \cite{DeRomeri:2022twg}.} 

\subsubsection{COHERENT Ge detector at the SNS}

\begin{figure}[t]
    \centering
    \includegraphics[scale=0.5]{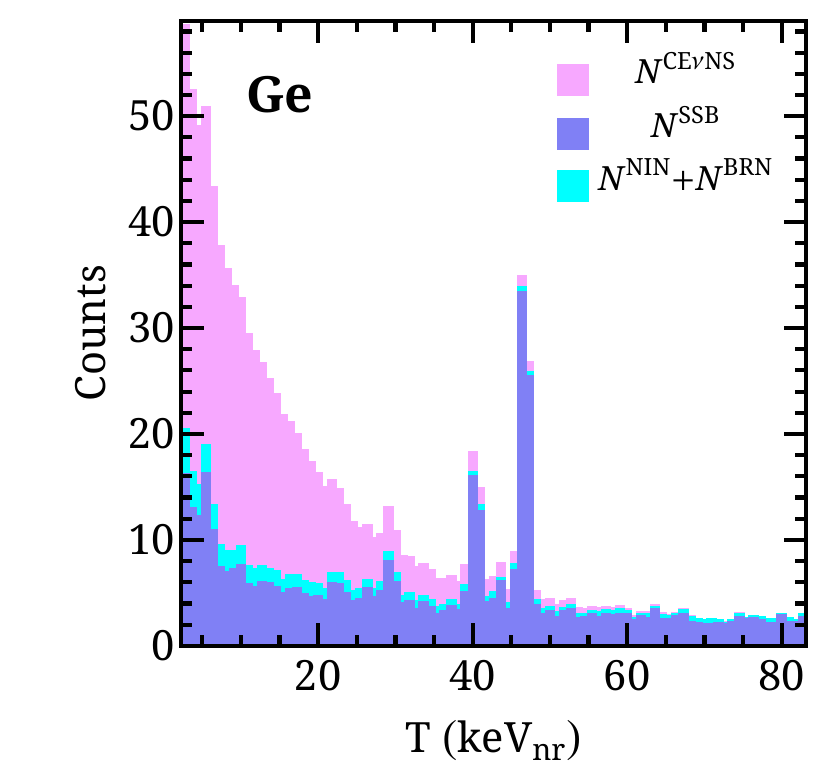}
    \includegraphics[scale=0.5]{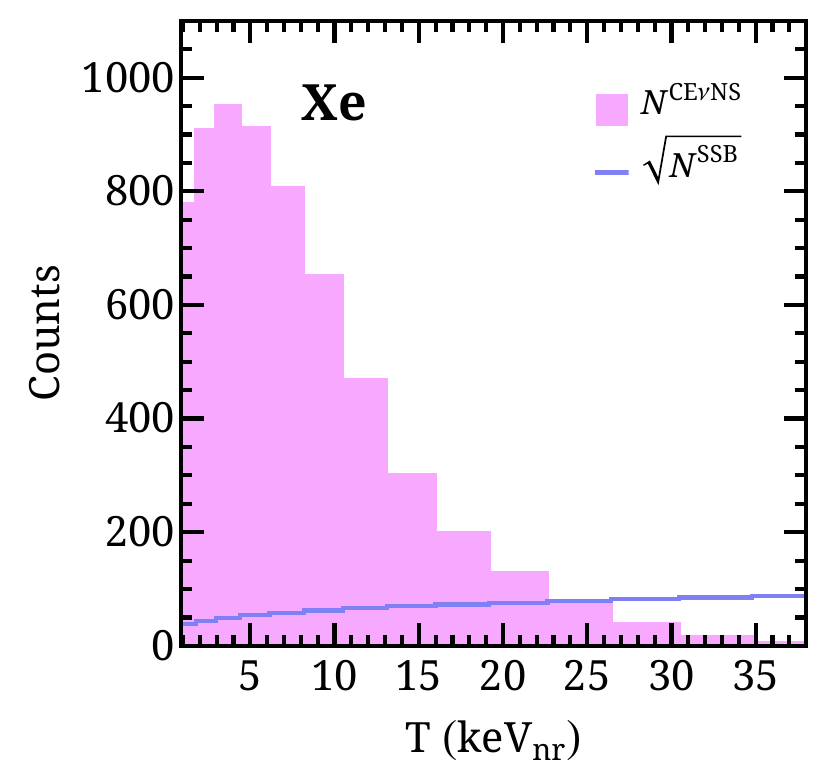}
    \caption{Expected number of events after one year of data taking at the  $\pi$-DAR source experiments under consideration. The left panel corresponds to the signal (CE$\nu$NS + backgrounds) predicted for the Ge detector experimental setup, while the right panel presents the spectrum expected for the xenon detector at the ESS described in the text. For clarity, the blue line corresponds to the square root of the expected SSB backgrounds, denoted as $\sqrt{N^{\textrm{SSB}}}$.}
    \label{fig:events:pi:dar}
\end{figure}

So far, the COHERENT collaboration has reported CE$\nu$NS measurements with two different detectors, one based on CsI, as discussed above, and another one using LAr. 
However, the complete experimental program of COHERENT  includes two other detectors:  COH-Ge-1 and  COH-NaI-2~\cite{Akimov:2022oyb}. 
Here we will focus on the expected sensitivity from COH-Ge-1,  a germanium-based detector array that, when completed, will consist of  eight p-type point-contact Ge
detectors with an approximate mass of 2 kg each, giving a total mass of around 16 kg of detector material, located at 22 m from the neutrino source \cite{Akimov:2022oyb}.
Regarding SNS characteristics, this facility currently operates at a proton beam energy of 1 GeV, with a power beam of 1.4 MW, although it is expected to be increased  up to 2 MW in the near future.
For our simulation, we  consider this last value of the beam power, corresponding to $N_{\textrm{POT}} = 2.24\times10^{23}$ for one calendar year of SNS operations, and to approximately 5000 hours of data taking~\cite{Akimov:2022oyb}.
For the number of neutrinos per flavor released,  we take the current reported value of $r = 0.0848$ \cite{COHERENT:2021xmm}. 
Regarding the energy resolution of the detector, we consider a Gaussian smearing function with  85 keV$_{ee}$ FWHM, and  we assume a conservative efficiency of 95$\%$. 
A summary of all the parameters considered for the calculation of the number of events for this experiment is given in Tab. \ref{tab:snschar}.
With all these considerations, we simulate the spectrum for one calendar year of SNS operations, as shown in the left panel of Fig. \ref{fig:events:pi:dar}, where the bin size corresponds to 1 keV$_{\textrm{ee}}$ assuming a constant QF of 0.22.
Besides the spectrum of CE$\nu$NS events, we also include in the figure the three main expected sources of backgrounds: SSB, and NIN+BRN.
The distribution of background events, taken from \cite{fnalNeutrino2020},
 shows that  SSB constitutes the dominant contribution for backgrounds at the SNS.

\begin{table}[b]
\centering
\begin{tabular}{ccccccc}
\toprule
\textbf{~~Experiment~~} & \textbf{~~Mass (kg)~~} & \textbf{~~Distance (m)~~} & \textbf{~~r~~} & \textbf{~~$N_{\textrm{POT}}$} ($\times10^{23}$) ~~ & \textbf{~~$T_{Th}$(keV$_{nr}$)~~} \\ 
\toprule
CsI            & 14.3                 & 19.3                    & 0.0848     & 3.198                &  4.18                               \\
\hline
Ge            & 16                 & 22                    & 0.0848     &  2.24                & 2.272                               \\
\hline
Xe                  & 20                 & 20                    & 0.3        & 2.8                 & 0.9     \\
\toprule
\end{tabular}
\caption{Relevant parameters for the simulation of the CE$\nu$NS signal at $\pi$-DAR source experiments.}
\label{tab:snschar}
\end{table}

\subsubsection{Xenon detector at the ESS }

Here we will also analyze the expected sensitivity of a CE$\nu$NS detector located at the future ESS.
Even though this facility is still under construction, several experimental proposals are considering  its particle physics potential \cite{Abele:2022iml}, including the possibility of measuring CE$\nu$NS with different detection materials. 
As a working example, we consider the case of a Xe-based detector with the same characteristics as proposed in \cite{Baxter:2019mcx} and summarized in Tab.~\ref{tab:snschar}. 
One of the main advantages of using the ESS is its relatively large maximum beam energy of 2 GeV and its beam power of 5 MW, under which $r = 0.3$. This allows for having $N_{\textrm{POT}} = 2.8\times10^{23}$ in a calendar year of operation, which, as in the case of the SNS, corresponds to approximately 5000 hours. 
 On the negative side, however, one finds that the larger frequency pulse of ESS reduces the background discrimination power of the experiment.
As a result, SSB contributions, reported in \cite{Baxter:2019mcx}, are larger than in the SNS. In contrast, the contributions from BRN and NIN in this experiment are expected to be negligible~\cite{Baxter:2019mcx}.

The relevant parameters necessary to calculate the expected number of CE$\nu$NS events are shown in Tab.~\ref{tab:snschar}. Regarding detector characteristics, we considered a Gaussian smearing function and an efficiency of 80$\%$ following the prescription given in \cite{Baxter:2019mcx}. The obtained spectrum and modeled backgrounds are shown in the right panel of Fig. \ref{fig:events:pi:dar}. Our result is consistent with the spectrum reported in Ref.~\cite{chatterjee2023constraining}, where they consider a three-year data taking for the same detector characteristics.

\subsection{Reactor experiments}

As discussed in the introduction,  there is great interest within the community to pursue new CE$\nu$NS measurements from neutrino sources other than  $\pi$-DAR. 
Nuclear reactors can constitute a very interesting source for CE$\nu$NS neutrino experiments. There,  electron antineutrinos are emitted by the beta decay of neutron-rich elements produced in the chain reactions that take place inside the reactor.
The main contribution to this neutrino flux comes from the decay of four elements with different average proportions, which depend on the characteristics of the nuclear reactor~\cite{Kopeikin:2012zz}. Here we will assume $^{235}$U: 0.56, $^{239}$Pu: 0.31, $^{238}$U: 0.07, $^{241}$Pu: 0.06.
These fission fractions vary over time as the fuel inside the reactor burns, so one can consider an average value. 
For energies above 1.8 MeV, where kinematics allows the detection of reactor antineutrinos via inverse beta decay, the antineutrino spectrum can be modeled by the Huber-Mueller parametrization~\cite{PhysRevC.84.024617, Mueller:2011nm}
\begin{equation}
    \lambda(E_\nu) = \sum_{m} f_m \exp\left [  {\sum_{k = 1}^{6}\alpha_{m,k}E_\nu^{k-1}} \right ] \,,
   \label{eq:reac:flux}
\end{equation}
where $f_m$ are the fission fractions discussed above, and the coefficients $\alpha_{m,k}$ are defined in Refs.~\cite{PhysRevC.84.024617,Mueller:2011nm}.
Below the kinematic threshold for inverse beta decay, the antineutrino spectrum can be modeled as in Ref. \cite{Kopeikin:2004cn}. However, for the thresholds in CE$\nu$NS experiments, only neutrino energies above 2 MeV are relevant. 
 The total number of events for a given reactor experiment is determined from Eq.~(\ref{eq:Nevents_CEvNS_alpha}), by substituting the $\pi$-DAR neutrino spectrum by the normalized reactor spectrum in Eq.~(\ref{eq:reac:flux}) multiplied by the corresponding flux. 

 Following the description above, we have estimated the sensitivities of possible future CE$\nu$NS searches at reactor neutrino experiments. The proposed experimental setups are based on the reported configuration for backgrounds, thresholds, and masses of the CONUS~\cite{CONUS:2020skt} and $\nu$GeN~\cite{nGeN:2022uje} experiments.
One of the main advantages of studying CE$\nu$NS with reactor neutrinos is that, due to their energy, the expected signal is blind to the effects of the proton and neutron form factors. 
This is not the case for stopped pion neutrinos, where the role of form factors is relevant and can  be confused with other new physics contributions to the CE$\nu$NS cross section. 
Therefore, the combined analysis of CE$\nu$NS experiments using $\pi$-DAR and reactor neutrino sources can help disentangle the nuclear physics parameters from the potential signals of physics beyond the SM.

\subsubsection{CONUS-like Ge detector}
\begin{figure}[t]
    \centering
    \includegraphics[scale=0.5]{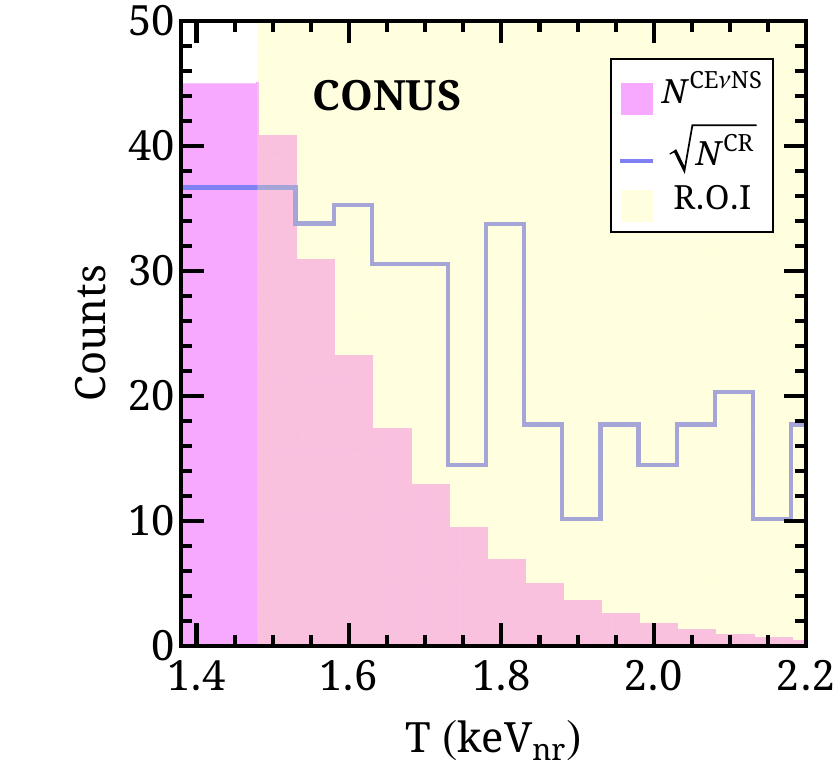}
    \includegraphics[scale=0.5]{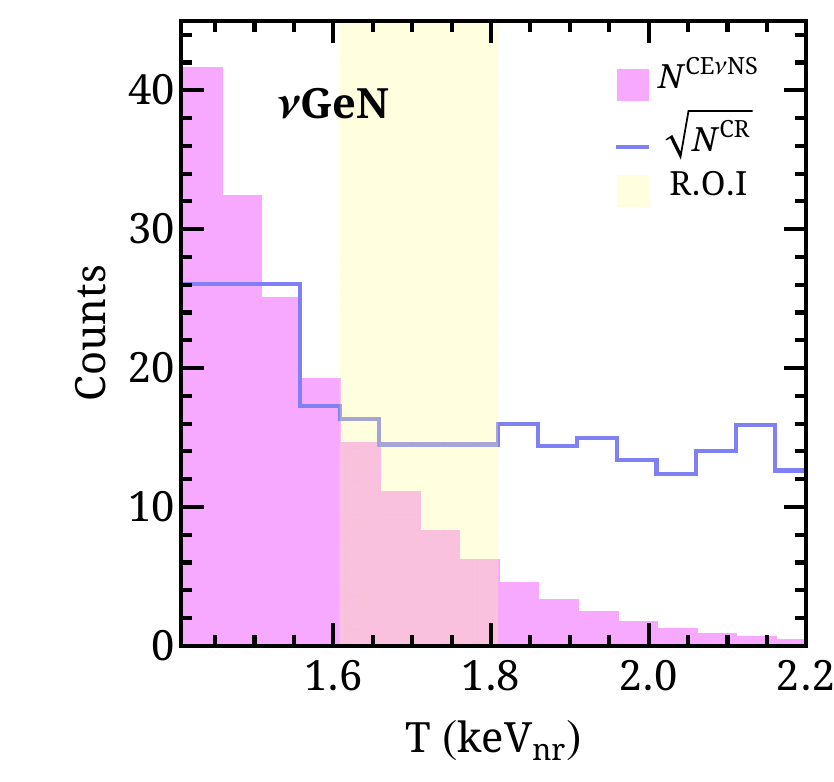}
    \caption{
    Expected CE$\nu$NS spectrum at reactor experiments. Left panel shows the expected signal at the considered CONUS-like experiment described in the text. Right panel shows the corresponding expected events for the $\nu$GeN-like experiment. In both cases, the squared root of the cosmic ray backgrounds, $N^{\textrm{CR}}$, is shown in blue and the experimental region of interest is indicated between yellow bands.}
    \label{fig:events:reactor}
\end{figure}

CONUS was a CE$\nu$NS experiment having the reactor at the Brokdorf power plant (KBR) as its neutrino source~\cite{Buck:2020opf}, with a thermal power of 3.9 GW and a total flux of $2.3\times 10^{13}$ antineutrinos/cm$^{2}$/s for an average distance of 17~m from the detector to the reactor. 
The CONUS collaboration has not observed a positive signal of CE$\nu$NS but has reported upper limits to the associated cross section \cite{CONUS:2020skt}. The upgrade of the detector to CONUS$+$ is currently underway~\cite{magcevns:edgar}.
Here we study the sensitivity that a detector similar to  CONUS can reach in a scenario where the measurement of CE$\nu$NS is achieved. During its first two runs,  CONUS  used four high-purity p-type Ge detectors, each with an approximate mass of 1~kg. As reported in Ref.~\cite{CONUS:2020skt}, each detector had different thresholds and regions of interest for data analysis. To study the future sensitivity of a CONUS-like experiment, we will assume four detectors with the same characteristics as the detector C1 in Ref.~\cite{CONUS:2020skt}, for one year of data taking. This means that we assume 4 kg of total fiducial mass and a threshold of 0.296 keV$_{\textrm{ee}}$. 
 A summary of the relevant parameters for the calculation of the total number of events is given in Tab.~\ref{tab:reac:char} (see Sec.~\ref{sec:NSI:reac1}). The background model for the analysis is taken from Ref.~\cite{CONUS:2020skt}, and scaled to one year of data taking. Note that the dominant contribution to backgrounds comes from cosmic rays and, then, it is not correlated with reactor backgrounds, which are negligible~\cite{CONUS:2021dwh}. Note as well that the background spectrum is presented in bins of width 0.01 keV$_{\textrm{ee}}$. 
 To convert this signal into the equivalent one in nuclear recoil energy, we assume an average QF of 0.19 in the region of interest. As for the smearing function, by following the prescription in \cite{CONUS:2020skt}, we assume a detector efficiency of 95\% and a Gaussian smearing with 85 keV$_{ee}$ FWHM. Under these assumptions, we get the distribution shown in the left panel of Fig.~\ref{fig:events:reactor}, where we have considered a $^{72}$Ge detector. For reference, the yellow band in the figure indicates the region of interest for the C1 detector, conveniently chosen by the collaboration to increase the efficiency and reduce backgrounds~\cite{CONUS:2020skt}.

\subsubsection{$\nu$GeN-like Ge detector}

As a second working example of reactor sources, we consider the characteristics of the $\nu$GeN experiment~\cite{nGeN:2022uje}. 
As in the case of CONUS, a signal of CE$\nu$NS has not been reported by the $\nu$GeN Collaboration yet. 
Then, here we will analyze the potential sensitivity of the $\nu$GeN-like experimental setup assuming a positive measurement can be achieved. 
Currently, the $\nu$GeN experiment is located at the Kalinin nuclear power plant (KNPP), with a thermal power of 3.1 GW and an impressive flux of $4\times 10^{13}$ antineutrinos/cm$^{2}$/s  for a detector located 11 m from the core of the reactor. 
The experiment consists of a p-type Ge detector with a current fiducial mass of 1.4 kg, that can be further extended. As indicated in Ref.~\cite{nGeN:2022uje}, the region of interest for this detector goes from 0.32 to 0.36 keV$_{\textrm{ee}}$. To convert those energies into nuclear recoil equivalent energies,  we assume a constant QF of 0.20. A summary of the detector characteristics is given in Tab.~\ref{tab:reac:char} (see Sec. \ref{sec:NSI:reac1}). Regarding the smearing, we assume a Gaussian distribution with 10.4~keV$_{\textrm{ee}}$
FWHM, and we take the efficiency from \cite{nGeN:2022uje}. Under these assumptions, we simulate the expected event rate shown in the right panel of Fig.~\ref{fig:events:reactor}, where we also present the backgrounds taken from Ref.~\cite{nGeN:2022uje} and scaled to one-year data taking, again with a dominant contribution from cosmic rays. Yellow bands in the figure indicate the region of interest reported by the collaboration. 

\section{Analysis}\label{sec:analysis}

Once we have presented the different experimental setups under consideration, we now describe the strategy used to test the neutron rms radius as well as to constrain NSI parameters. Our ultimate goal will be combining different neutrino sources to constrain these parameters when simultaneously studied. 
The procedure we follow is essentially based on Ref.~\cite{canas2020interplay}, although here we consider additional NSI parameters and present our results with the up-to-date data from CsI COHERENT detector. Besides, we consider more realistic scenarios of backgrounds and efficiencies regarding future experiments.

 We first analyze the current data from the CsI detector at the SNS using the following Poissonian $\chi^2$  function~\cite{DeRomeri:2022twg}
\begin{equation}
  \chi^2 = 2\sum_{i,j}\left [ N^{\textrm{th}}_{ij}(X) - N^{\textrm{exp}}_{ij} + N^{\textrm{exp}}_{ij}\ln\left ( \frac{N^{\textrm{exp}}_{ij}}{N^{\textrm{th}}_{ij}(X)} \right )\right ] + \sum_{m=1}^2 \frac{\alpha_{m}}{\sigma_{\alpha_m}^2} + \sum_{k=1}^3\frac{\beta_k}{\sigma_{\beta_k}^2} ,
    \label{eq:chi:CsI}
\end{equation}
with
\begin{equation}
    N^{\textrm{th}}_{ij}(X) = (1+\alpha_1) N^{\textrm{CE$\nu$NS}}_{ij}(X,\alpha_2,\alpha_3) +  (1+\beta_1)N^{\textrm{SSB}}_{ij}+  (1+\beta_2)N^{\textrm{BRN}}_{ij}(\alpha_3)+  (1+\beta_3)N^{\textrm{NIN}}_{ij}(\alpha_3) \,.
    \label{N:chi:CsI}
\end{equation}
Here, the $i$ index runs over recoil energy bins and the $j$ index runs over the time bins, being $N_{ij}^{\textrm{exp}}$ the measured number of events at each bin \footnote{See Ref. \cite{Pershey:M7s} and Appendix A of Ref. \cite{DeRomeri:2022twg}}, and $N_{ij}^{\textrm{th}}(X)$ the predicted total number of events, which includes CE$\nu$NS interactions (as a function of a set of the parameters under test, $X$), as well as the contributions from SSB, BRN, and NIN background events, denoted as $N_{ij}^{\textrm{SSB}}$, $N_{ij}^{\textrm{BRN}}$, and $N_{ij}^{\textrm{NIN}}$, respectively.
The parameters $\alpha_m$ and $\beta_m$ in Eqs.~(\ref{eq:chi:CsI}) and (\ref{N:chi:CsI}) are nuisance parameters with respect to which we perform the minimization process, each with its associated systematic uncertainty $\sigma_{\alpha_m}$ and $\sigma_{\beta_k}$, which values are taken from \cite{DeRomeri:2022twg}. 
The parameters indicated as $\alpha_m$ have an impact on the CE$\nu$NS signal and background predictions, $\alpha_1$ being solely associated with the flux normalization and QF. On the other hand, the parameter $\alpha_2$ is introduced as in \cite{DeRomeri:2022twg} to account for the CE$\nu$NS efficiency uncertainty, and $\alpha_3$ for the uncertainty in timing distribution, which also has an effect on the BRN and NIN background distributions. 
In addition, $\beta_k$ parameters are introduced to account for background normalizations, with $k = 1, 2, 3$ for SSB, BRN, and NIN, respectively. 

In the case of future experiments, we do not consider timing distribution, and we perform the analysis by minimizing the function
\begin{equation}
  \chi^2 = 2\sum_{i}\left [ N^{\textrm{th}}_{i}(X) - N^{\textrm{exp}}_{i} + N^{\textrm{exp}}_{i}\ln\left ( \frac{N^{\textrm{exp}}_{i}}{N^{\textrm{th}}_{i}(X)} \right )\right ] + \frac{\alpha_1}{\sigma_{\alpha_1}^2} + \sum_{k}\frac{\beta_k}{\sigma_{\beta_k}^2}.
    \label{eq:chi:future}
\end{equation}
with
\begin{equation}
    N^{\textrm{th}}_i(X) = (1+\alpha_1) N^{\textrm{CE$\nu$NS}}_i(X) + \sum_{k} (1+\beta_k)N^{\textrm{bckg-}k}_i.
    \label{N:chi:future}
\end{equation}
where the $i$ index runs over recoil energy bins. In the previous equation,  $N_{i}^{\textrm{th}}(X)$ is the theoretical number of events expected at each bin by considering the contribution from CE$\nu$NS, $N_{i}^{\textrm{CE$\nu$NS}}(X)$, and the corresponding backgrounds $N^{\textrm{bckg-}k}_i$. The parameter $\alpha_1$ has the same meaning as in Eq.~(\ref{N:chi:CsI}), and the parameters $\beta_k$ are associated with background normalizations. The nature and number of the different background contributions will be different depending on the experiment under analysis. 
In the case of a Ge detector at the SNS, we consider two nuisance parameters ($k = 1,2$); one denoted as $\beta_1$,  associated to SSB events ($N^{\textrm{SSB}}_i\equiv N^{\textrm{bckg}-1}_i$), and another one, denoted as $\beta_2$, for BRN+NIN events ($N^{\textrm{BRN+NIN}}_i\equiv N^{\textrm{bckg}-2}_i$). On the other hand, for the Xe detector at the ESS, we follow the procedure in Ref. \cite{Baxter:2019mcx}, and we neglect the BRN+NIN contribution to the number of events. For reactors, we only consider one dominant contribution to backgrounds coming from cosmic ray muons ($N^{\textrm{CR}}_i\equiv N^{\textrm{bckg}-1}_i$),  and hence one single nuisance parameter, $\beta_1$. 
For both neutrino source experiments, $N_i^{\textrm{exp}}$ refers to the measured number of events, which, as we are considering future experiments, we will assume as the SM prediction plus background contributions as described in the previous section. Following this procedure we will be able to estimate the sensitivity of a given experiment to the parameter under study through its impact on the CE$\nu$NS predicted signal.

\section{Results} \label{sec:results}

In this section, we present our main results for the sensitivity of the current and future discussed experiments to both the neutron rms radius and NSI parameters. We begin by studying the parameters individually, then we simultaneously study both parameters on $\pi$-DAR sources, which are sensitive to the neutron rms radius, and finally, we combine the results of the two different sources.

\subsection{Nuclear Physics from $\pi-$DAR experiments}
\label{sec:res:nucl}
\begin{figure}[t]
    \centering
    \includegraphics[scale = 0.5] {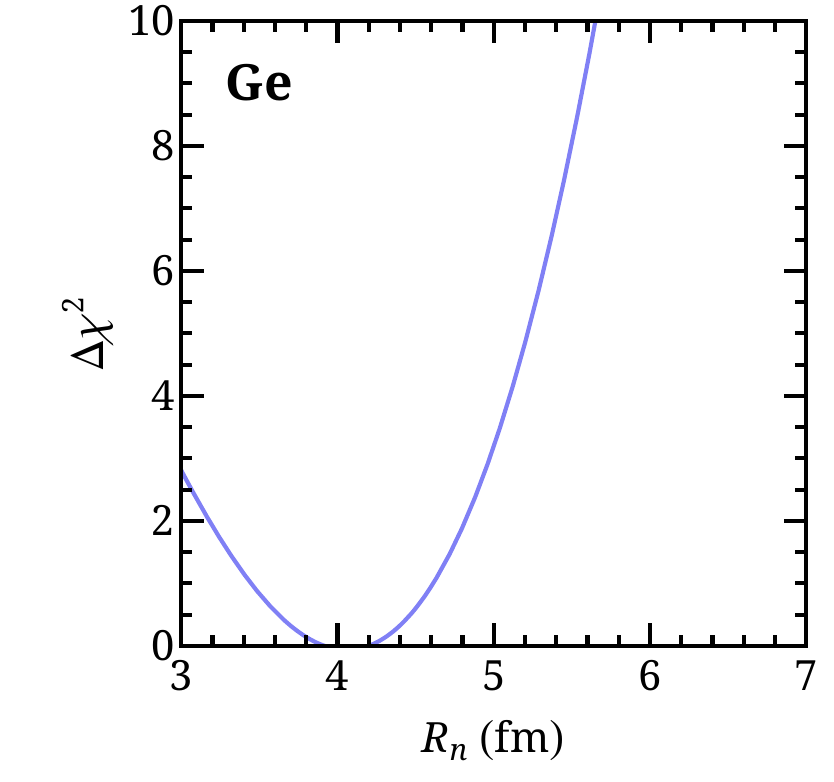}
    \includegraphics[scale = 0.5]
    {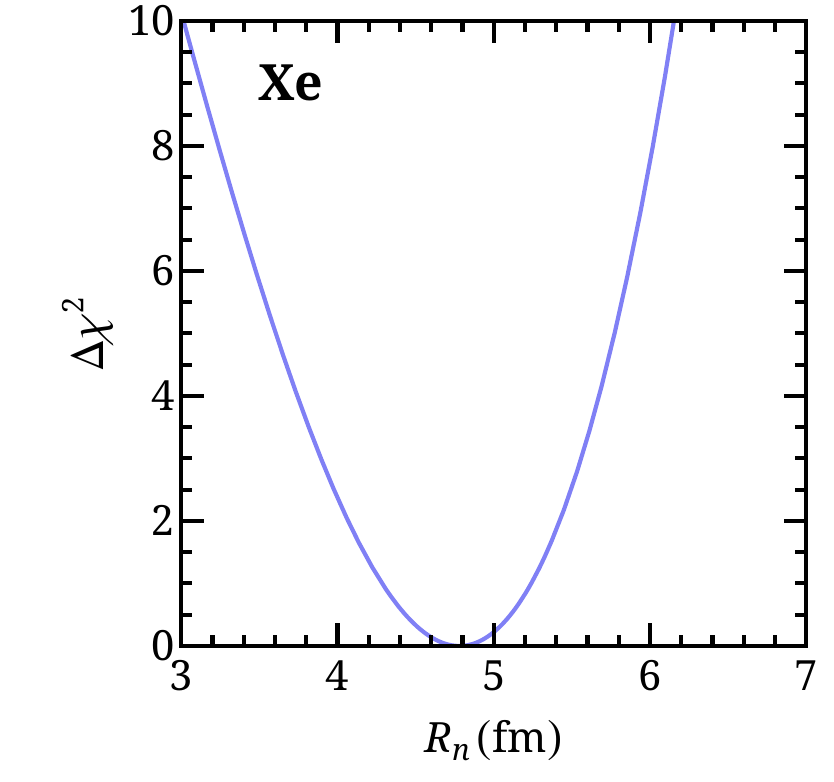}
    \caption{Expected sensitivity of $\pi$-DAR experiments to the neutron rms radius. The left panel shows the results for a Ge detector at the SNS and the right panel assumes a Xe detector located at the ESS. In both cases, one year of operations has been assumed.}
    \label{fig:rms1D}
\end{figure}

As discussed in the introduction, the process of CE$\nu$NS is sensitive to the neutron rms radius of the target material through the form factor defined in Eq.~(\ref{eq:KNFF}). Experimentally, this parameter has only been measured for some materials like $^{208}$Pb and $^{48}$Ca, which were determined from parity-violating processes at the PREX~\cite{Horowitz:2012tj} and CREX~\cite{CREX:2022kgg} experiments, respectively. A precise determination of this parameter for the target materials considered here
is of particular interest for neutrino physics. For instance, Ge detectors are used in the searches for neutrinoless double beta decay, and Xe detectors are also used in dark matter searches, where the CE$\nu$NS signal is part of the backgrounds that form the neutrino floor.

To test the sensitivity of $\pi$-DAR experiments to the neutron rms radius, we consider the cross section introduced in Eq. (\ref{eq:cross}), and we compute the number of events by varying the parameter $R_n$ in Eq. (\ref{eq:rms:n}) while keeping the proton rms radius, $R_p$, to a fixed value. An analysis involving the $R_{n}$ parameter alone for CsI data has already been discussed in Ref. \cite{DeRomeri:2022twg}, so we start our discussion with the expectations for the future proposals introduced in Sec. \ref{sec:source}. 

We first consider a $^{72}$Ge detector at the SNS with the characteristics described in Sec.~\ref{sec:source} for one calendar year of SNS operations, and with $R_p = 4.06$ fm \cite{Angeli:2013epw}. 
The results of the analysis are shown in the left panel of Fig.~\ref{fig:rms1D}. For the analysis, we have considered a conservative systematic error uncertainty of $\sigma_{\alpha_1} = 10$\%. In addition, we considered the SSB and BRN+NIN backgrounds as discussed in Sec. \ref{sec:source}, with their corresponding uncertainties of $\sigma_{\beta_1} = 5\%$ and $\sigma_{\beta_2} = 25\%$, which were motivated by the current status of the CsI and LAr detectors used by the COHERENT collaboration. 
The 90\% C.L. expected sensitivity of the experiment is given by
\begin{equation}
3.0~\textrm{fm} \leq R_{n}^{\textrm{Ge}} \leq 4.9~\textrm{fm} \,.
\end{equation}

As a second scenario, we study the expected sensitivity to the neutron rms radius of a Xe detector at the ESS with the characteristics described in Sec. \ref{sec:source}. Again, for the analysis, we vary the corresponding value of $R_n$ while keeping the proton rms radius fixed to a value of $R_p = 4.73~\textrm{fm}$ \cite{Angeli:2013epw}. The result for one year of ESS operations is shown in the right panel of Fig. \ref{fig:rms1D}, 
where we have only considered the contribution to backgrounds from SSB as described before. 
Assuming the  corresponding uncertainties as $\sigma_{\alpha_1} = 10$\% and $\sigma_{\beta_1} = 1\%$ \cite{Baxter:2019mcx}, we obtain the following expected sensitivity at 90\% C.L.
\begin{equation}
3.9~\textrm{fm} \leq R_{n}^{\textrm{Xe}} \leq 5.5~\textrm{fm} \,.
\end{equation}
Notice that the two panels in Fig.~\ref{fig:rms1D} are not directly comparable since we are working with different nuclei as target materials used for detection. However, we note that the constraints for the Xe detector are slightly better, as a consequence of the larger statistics expected for this detector since the CE$\nu$NS cross section effectively scales as $N^2$.

\subsection{NSI from $\pi-$DAR experiments}
\label{sec:res:nsi:pidar}

\begin{figure}[t]
    \centering
    \includegraphics[scale = 0.5]{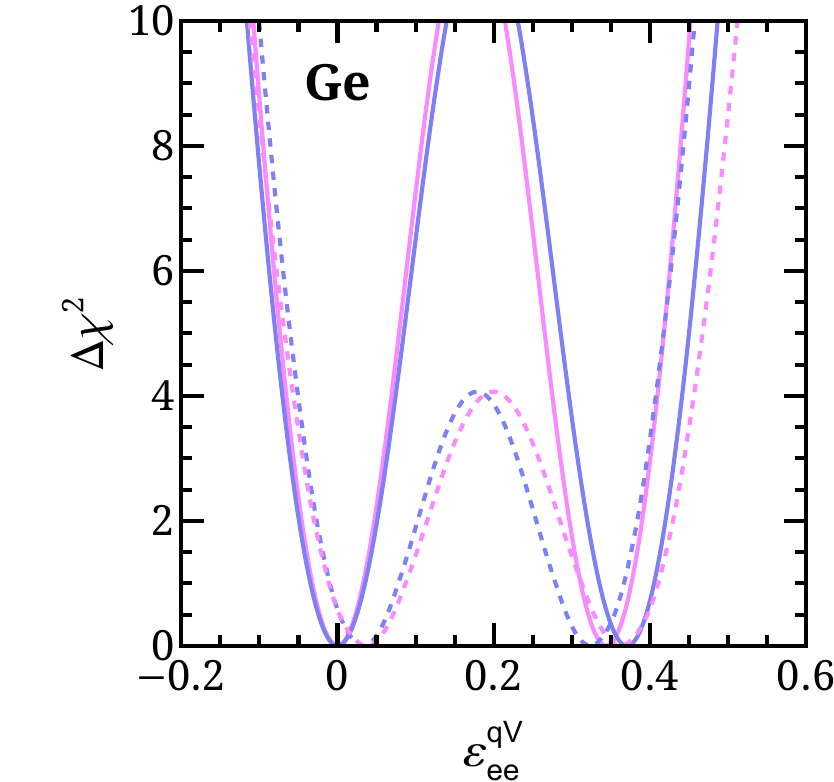}
    \includegraphics[scale = 0.5]{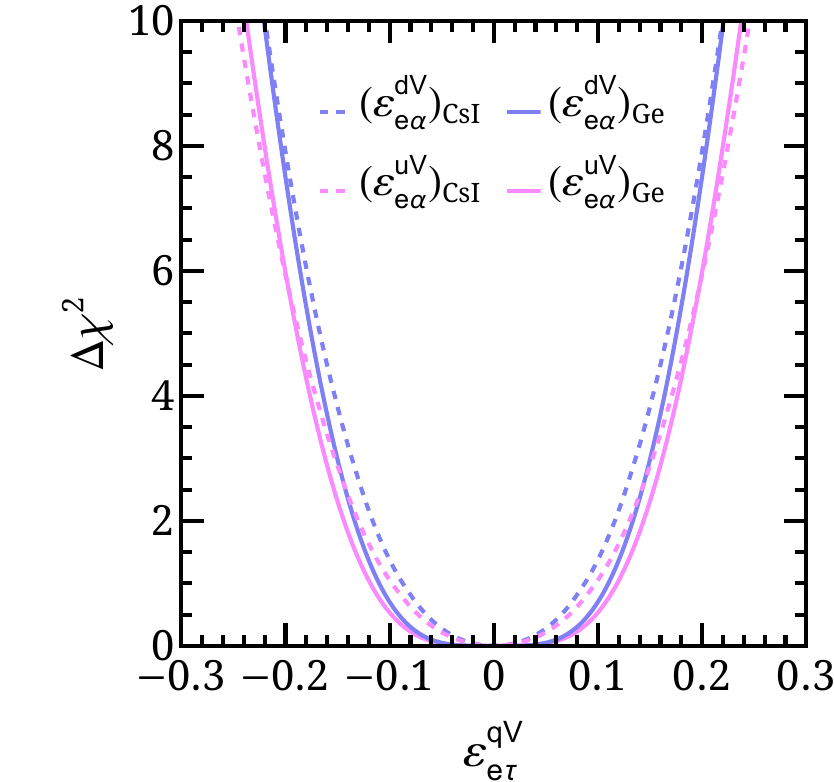}
    \caption{ Solid lines show the expected sensitivity of a Ge detector at the SNS to  non-universal (left panel) and flavor-changing (right panel) NSI couplings with $up$ (pink) and $down$ (blue) quarks. Dashed lines (with the same colour code) correspond to the current bounds obtained from the  analysis of COHERENT CsI data~\cite{DeRomeri:2022twg}.}
    \label{fig:NSI:DAR:SNS}
\end{figure}

\begin{figure}[t]
    \centering
    \includegraphics[scale = 0.5]{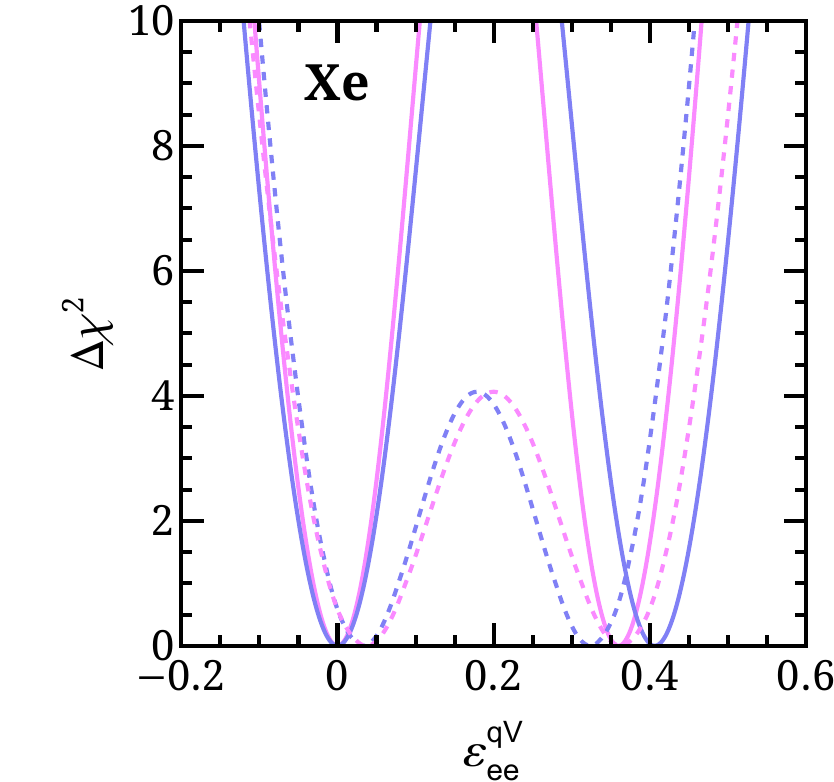}
    \includegraphics[scale = 0.5]{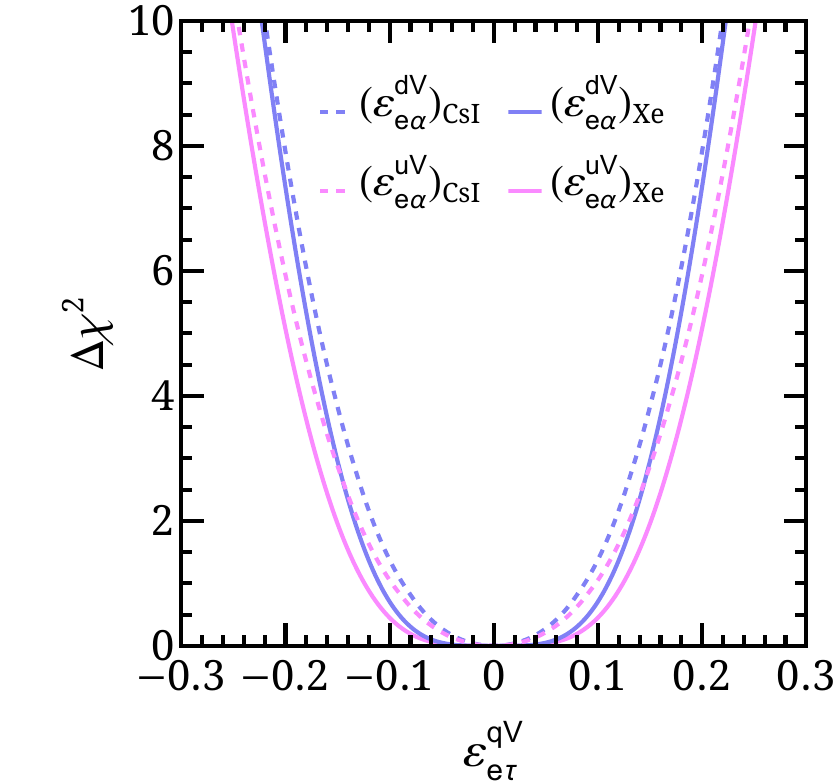}
    \caption{Solid lines show the expected sensitivity of a Xe detector at the EES to  non-universal (left panel) and flavor-changing (right panel) NSI couplings with $up$ (pink) and $down$ (blue) quarks. Dashed lines (with the same colour code) correspond to the current bounds obtained from the combined analysis of COHERENT data~\cite{DeRomeri:2022twg}. }
    \label{fig:NSI:DAR:ESS}
\end{figure}

We now turn our attention to the sensitivity of $\pi -$DAR source experiments to neutrino non-standard interactions at the detection point for the same experimental setups considered above. To this end, we now calculate the predicted number of events using the cross section with the weak charge as given in Eq.~(\ref{eq:weak_charge_NSIs}) and assuming only one NSI parameter to be different from zero at a time. 
This scenario has been discussed in detail for the CsI detector in Ref.~\cite{DeRomeri:2022twg} and here we present the analysis for the future proposals introduced in Sec.~\ref{sec:source}. 
The results for a $^{72}$Ge detector at the SNS are presented in Fig.~\ref{fig:NSI:DAR:SNS}, where we show the expected $\Delta\chi^2$ profiles for non-universal (left panel) and flavor changing (right panel) neutrino NSI  with $up$  (pink) and $down$ (blue) quarks. 

When compared to current bounds on NSI from CE$\nu$NS (see the dashed lines in the plot),  one sees that the expected constraints are slightly better than the ones obtained from the analysis of COHERENT measurements~\cite{DeRomeri:2022twg}, which is mainly driven by CsI data.
For the case of non-universal NSI in future experiments, we see a local minimum at $\varepsilon^{qV}_{ee}$ = 0, corresponding to the SM solution, and a second minimum at  $\varepsilon_{ee}^{uV} \approx 0.34$ and $\varepsilon_{ee}^{dV}\approx 0.37$. This is expected since, for these values, the combination of the standard and non-standard couplings mimics the SM prediction, as discussed in \cite{chatterjee2023constraining}. At 90\% C.L., we get the following expected sensitivity to  non-universal NSI
\begin{eqnarray}
    -0.058 \leq \varepsilon_{ee}^{dV} \leq 0.060 \quad & \cup &  \quad 0.310 \leq \varepsilon_{ee}^{dV} \leq 0.428 \, , \\
    -0.052 \leq \varepsilon_{ee}^{uV} \leq 0.055 \quad & \cup & \quad 0.288 \leq \varepsilon_{ee}^{uV} \leq 0.399\,.
\end{eqnarray}
Regarding flavor-changing NSI couplings, shown in the right panel of the same figure,  there is only one minimum, corresponding to $\varepsilon^{qV}_{e\tau} = 0$. This result is also expected since, in this case, there is no possible interference between the SM and the NSI contribution to the cross section, as seen from Eq.~(\ref{eq:weak_charge_NSIs}). The expected sensitivity at 90\% C.L. is 

\begin{equation}
|\varepsilon_{e\tau}^{dV}| \leq 0.145 \,, \quad |\varepsilon_{e\tau}^{uV}| \leq 0.156 \,.
\end{equation}
Notice that, in this case, the proposed future experiment would not be able to improve the current bounds on flavor-changing NSI for one year of data taking. However, in the next section, we will analyze the impact of combining these results with future reactor data.

For completeness, we show in Fig.~\ref{fig:NSI:DAR:ESS} the corresponding results for a potential Xe detector located at the ESS. As before, the left panel in the figure shows the results for non-universal NSI couplings, while the right panel presents the case of flavor-changing NSI.
As in the previous case, we find two solutions for the non-universal NSI couplings $\varepsilon_{ee}^{qV}$, although here the non-SM local minimum is shifted with respect to the one obtained for the Ge detector in Fig.~\ref{fig:NSI:DAR:SNS}.
This is because the position of this minimum depends on the ratio of protons to neutrons in the target material. 
The 90\% C.L. expected sensitivity of this experimental setup to $\varepsilon_{ee}^{qV}$ is given by
\begin{eqnarray}
    -0.058 \leq \varepsilon_{ee}^{dV} \leq 0.056 \quad & \cup & \quad 0.347 \leq \varepsilon_{ee}^{dV} \leq 0.465 \, , \\
   -0.052 \leq \varepsilon_{ee}^{uV} \leq 0.050 \quad & \cup & \quad 0.307 \leq \varepsilon_{ee}^{uV} \leq 0.411 \,,
\end{eqnarray}
while for flavor-changing NSI we  obtain
\begin{equation}
|\varepsilon_{e\tau}^{dV}| \leq 0.145 \quad \,, \quad |\varepsilon_{e\tau}^{uV}| \leq 0.165 \,.
\end{equation}

\subsection{NSI from reactor experiments}\label{sec:NSI:reac1}

\begin{figure}[t]
    \centering
    \includegraphics[scale=0.5]{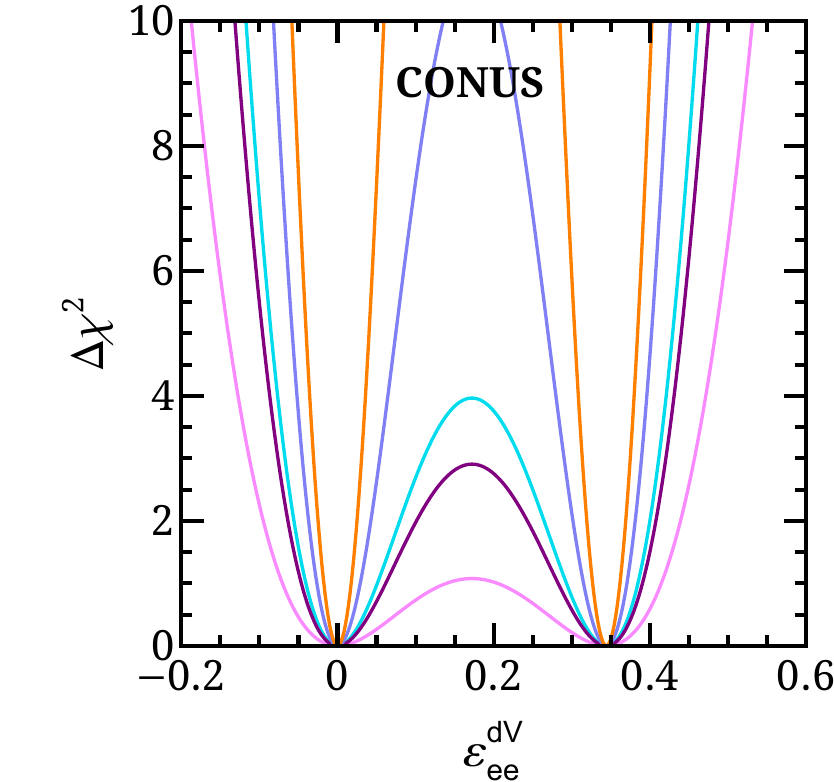}
    \includegraphics[scale=0.5]{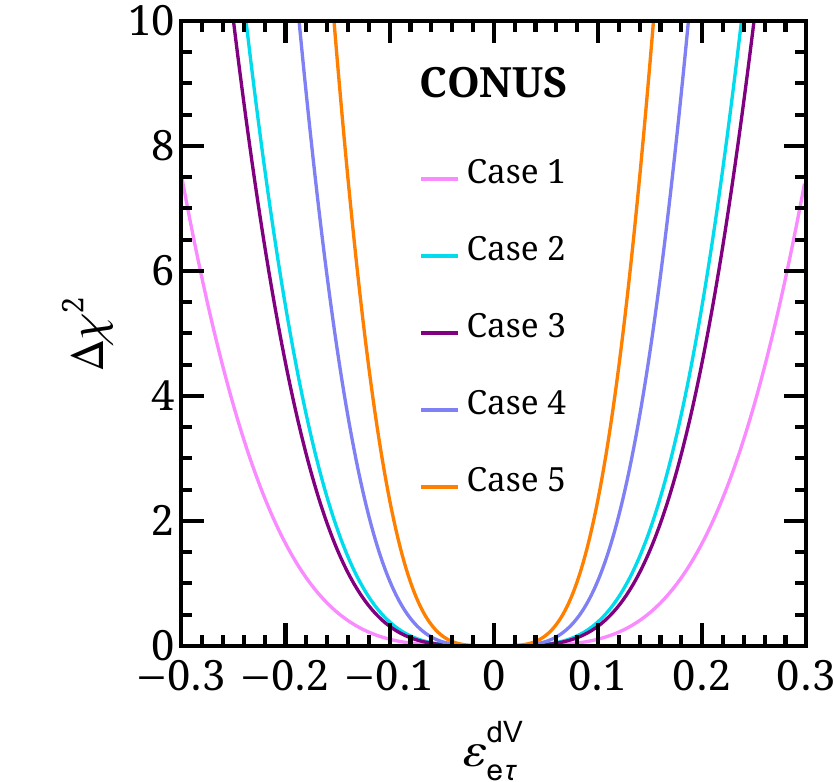}
    \caption{Expected sensitivity of a CONUS-like detector to non-universal (left panel) and flavor-changing (right panel) NSI couplings with $down$ quarks. The different configurations are defined in Table~\ref{tab:reac:char} and the color code is the same for the two panels.}
    \label{fig:reac:CONUS}
\end{figure}

\begin{figure}[t]
    \centering
    \includegraphics[scale=0.5]{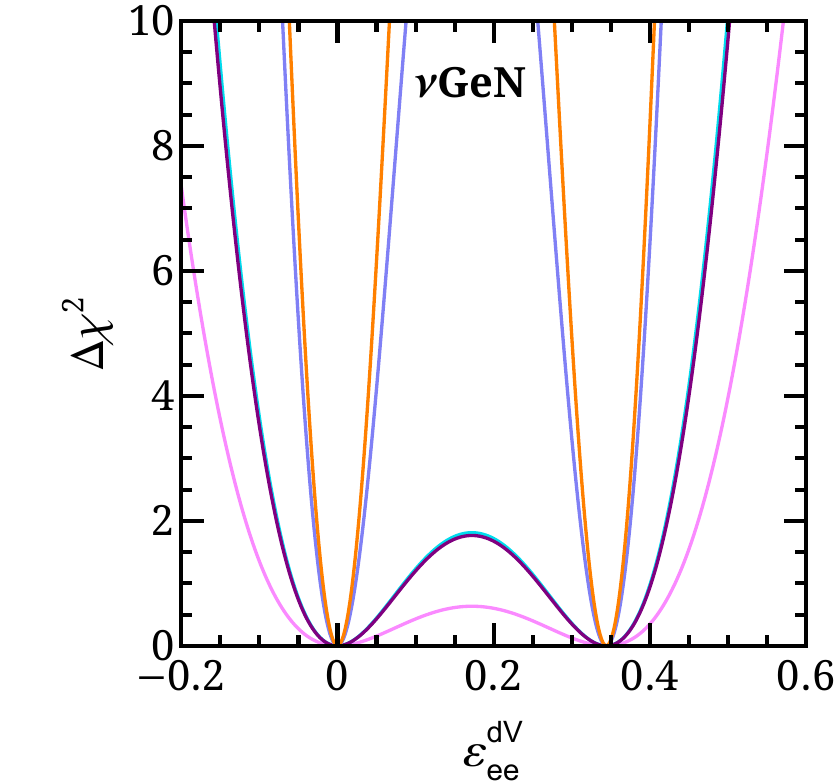}
        \includegraphics[scale=0.5]{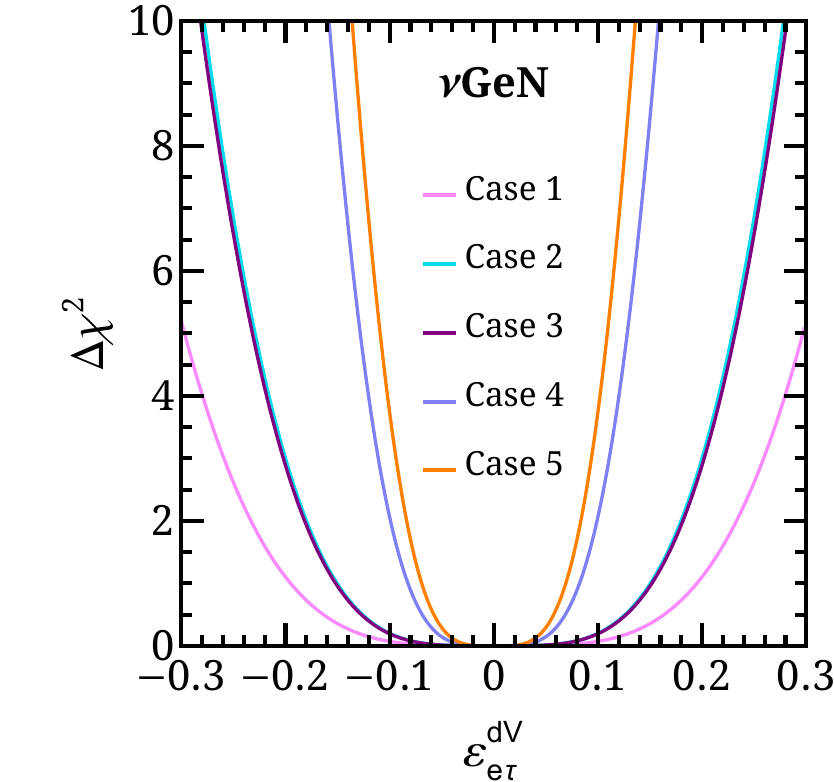}
    \caption{Same as Fig.~\ref{fig:eedV:rn:Ge:CONUS} for a $\nu$GeN-like detector for the experimental setups indicated in Tab.~\ref{tab:reac:char}. The color code is the same for the two panels.}
    \label{fig:reac:nugen}
\end{figure}

As we have already discussed, reactor neutrinos are not sensitive to the nuclear information of the target material. Therefore, in this section, we will only focus on the expected sensitivity of reactor experiments to NSI. 
We first consider a CONUS-like experiment with the background model and detector characteristics described in Sec.~\ref{sec:source}. Following the same procedure as in the SNS case, we assume only one NSI parameter to be different from zero at a time. The results are shown in Fig. \ref{fig:reac:CONUS} for non-universal (left panel) and flavor-changing (right panel) NSI. 
The pink lines (labeled as "Case 1" in the figure) correspond to the CONUS reported characteristics regarding mass and threshold, as discussed in Sec.~\ref{sec:source}. 
Since the results are qualitatively similar for $up$ and $down$ quark couplings (see Fig. \ref{fig:NSI:DAR:SNS}), here we focus on the sensitivities to NSI with $down$ quarks, exploring as well the  sensitivity of  eventual future scenarios for a CONUS-like experiment.
Thus, besides "Case 1", we consider four additional possibilities, allowing for larger detector masses and lower energy thresholds, see Tab.~\ref{tab:reac:char}.

Coming back to Fig.~\ref{fig:reac:CONUS}, the left panel shows that, for case 1, at  90\% C.L. the coupling $\varepsilon_{ee}^{dV}$ is constrained to a single interval that contains the two minima discussed before. 
By lowering the threshold, as in case 3,  the sensitivity is improved and the degeneracy breaks at 90\% C.L. This enhanced sensitivity can also be obtained by increasing the mass of the detector from the original CONUS configuration of 4 kg to 16 kg while keeping the original threshold, see case 2. 
Under these conditions, the degeneracy in the determination of the NSI parameter is clearly broken and now we have two separate intervals at a 90\% C.L. 
Another optimistic scenario is given in case 4, where both the mass and threshold are significantly improved with respect to case 1. 
Finally, case 5 represents the most optimistic situation in which we consider three years of data collection in addition to a larger mass and lower threshold. 
In the right panel of the figure, we show the corresponding results for the flavor-changing NSI parameter $\varepsilon_{e\tau}^{dV}$.  As expected, improving the mass and threshold of the detector results in a better sensitivity, getting the best constraint for case 5.

\begin{table}[b]
\centering
\begin{tabular}{cccclcccc}
\toprule
\multicolumn{4}{c}{\textbf{CONUS-like}}      &          & \multicolumn{4}{c}{\textbf{$\nu$GeN-like}}             \\ 
\multicolumn{4}{c}{d = 17 m, \, $\phi$ = 2.3$\times$10$^{13}$ cm$^{-2}$s$^{-1}$}      &          & \multicolumn{4}{c}{d = 11 m, \, $\phi$ = 4$\times$10$^{13}$ cm$^{-2}$s$^{-1}$}             \\ 
\hline
Case & Time (yr) & Mass (kg) & ROI (keV$_{ee}$)          &                     & Case & Time (yr) & Mass (kg) & ROI (keV$_{ee}$)            \\
\toprule
1    & 1         & 4         & {[}0.296,0.446{]}      &   ~~~~        & 1    & 1         & 1.4       & {[}0.32,0.36{]} \\
\hline
2    & 1         & 16        & {[}0.296,0.446{]}     &         & 2    & 1         & 16        & {[}0.32,0.36{]} \\
\hline
3    & 1         & 4         & {[}0.276,0.446{]}     &             & 3    & 1         & 1.4       & {[}0.30,0.45{]} \\
\hline
4    & 1         & 16        & {[}0.276,0.446{]}     &              & 4    & 1         & 16        & {[}0.28,0.45{]} \\
\hline
5    & 3         & 16        & {[}0.276,0.446{]}    &                & 5    & 3         & 16        & {[}0.28,0.45{]} \\ 
\toprule
\end{tabular}
\caption{Different cases analyzed for the CONUS-like and $\nu$GeN-like reactor experiments.}
\label{tab:reac:char}
\end{table}

As a second experimental scenario,  we consider a $\nu$GeN-like detector, with the same characteristics and background model as in Ref. \cite{nGeN:2022uje}. The sensitivity to neutrino NSI in this case is presented in Fig.~\ref{fig:reac:nugen}.
As before, solid pink lines represent the expected sensitivity for a detector with the same characteristics described in Sec. \ref{sec:source} and labeled as case 1. The other lines in the figure correspond to more optimistic scenarios that result from either increasing the detector mass or broadening the region of interest in nuclear recoil energy (indicated as the yellow region in the right panel of Fig.~\ref{fig:events:reactor}) with respect to case 1. The different experimental setups under consideration  are listed in Tab.~\ref{tab:reac:char}. Qualitatively, we see that these results are similar to the case of the CONUS-like detector.

\subsection{Constraining neutron rms radius and NSI simultaneously}

After analyzing nuclear effects and NSI individually, we now proceed to test these two physical cases simultaneously at $\pi$-DAR sources. 
The motivation behind this analysis lies in the fact that the uncertainties in nuclear form factors and the presence of neutrino NSI with matter might be confused in a $\pi$-DAR experiment. Thus, a discrepancy between the predicted and the observed signal in this type of experiment could be interpreted in terms of one of these hypotheses or even the combination of the two of them.  Interestingly, the joint analysis with CE$\nu$NS searches at reactor experiments, not sensitive to nuclear uncertainties, might help to lift the degeneracy between the two scenarios and improve the determination of nuclear and NSI parameters.

\begin{figure}[t]
    \centering
    \includegraphics[scale=0.35]{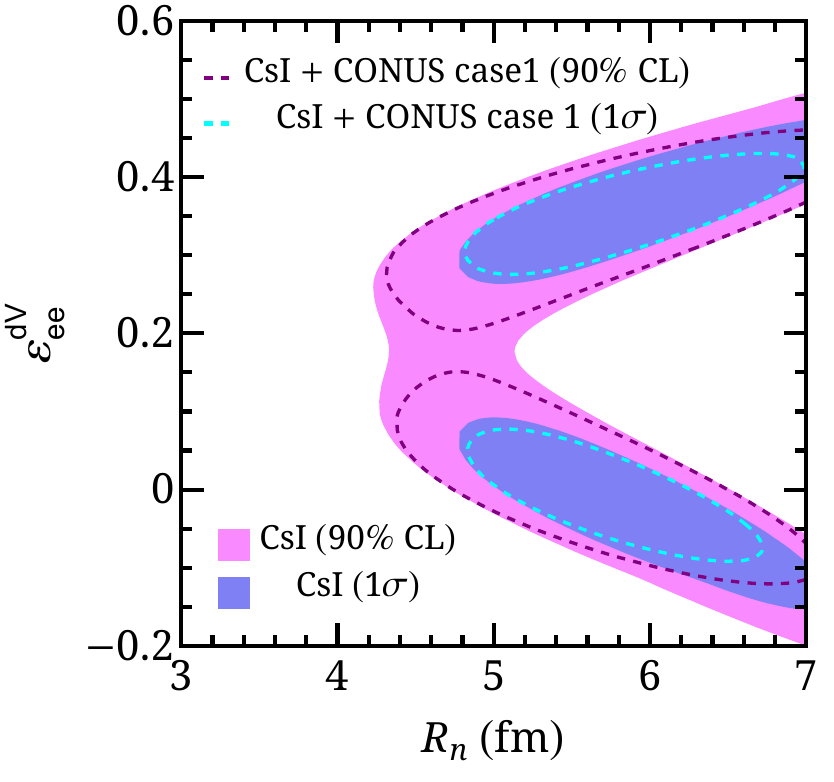}
    \includegraphics[scale=0.35]{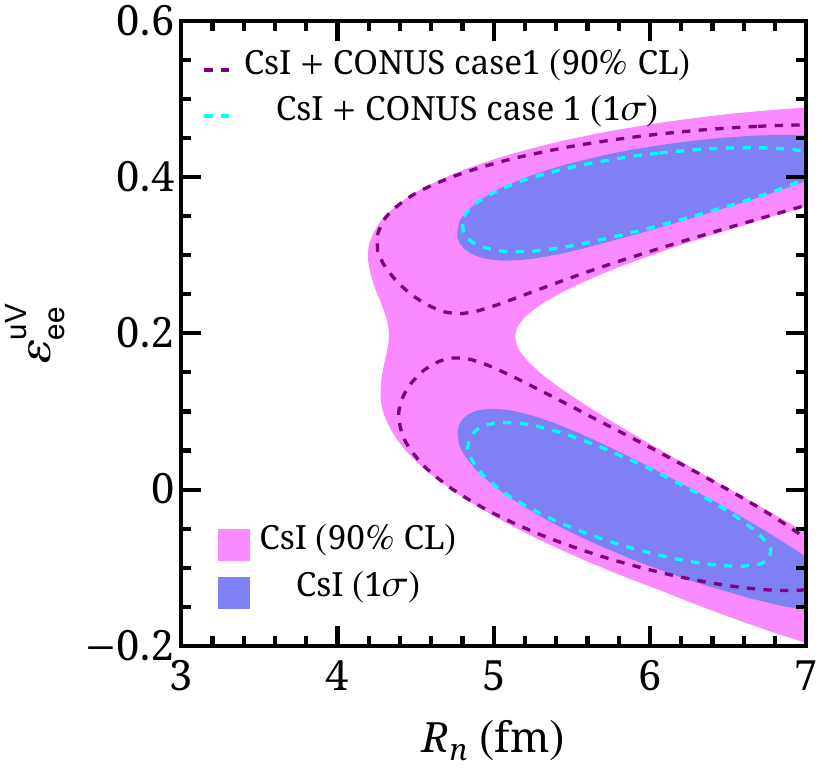}
    \includegraphics[scale=0.35]{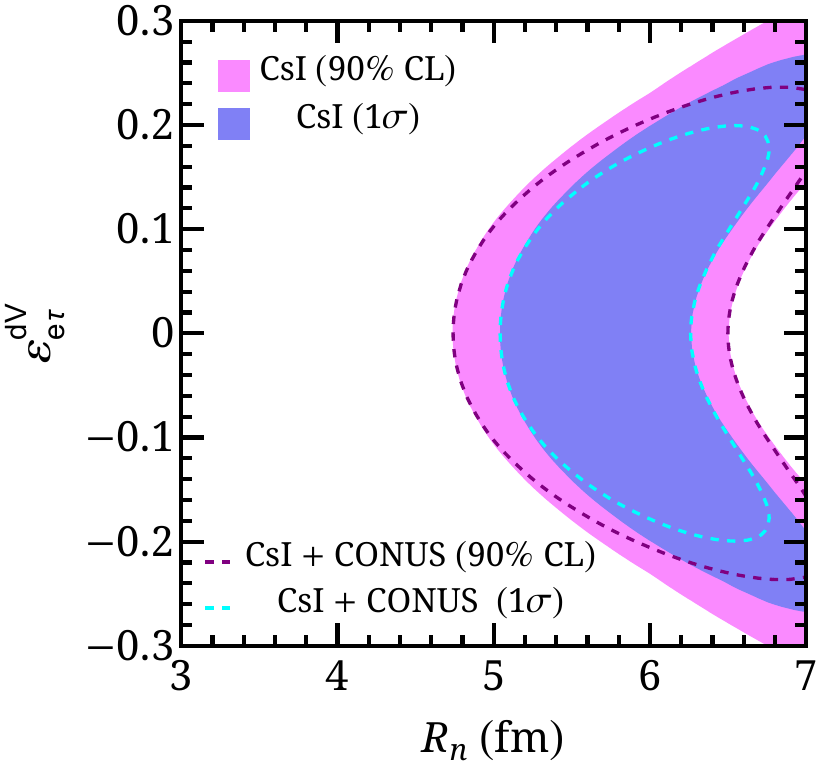}
     \includegraphics[scale=0.35]{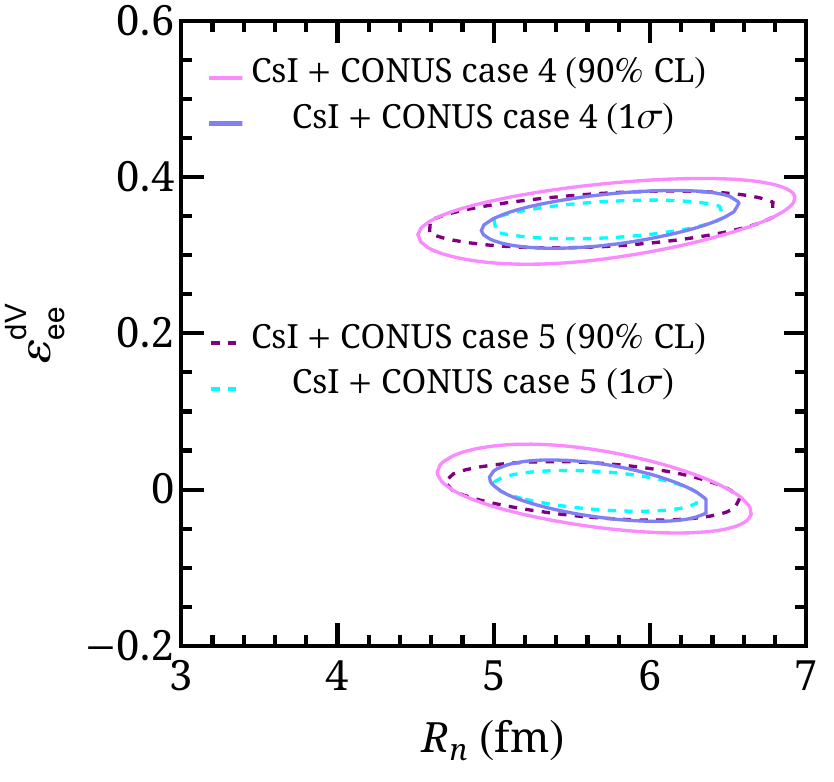}
     \includegraphics[scale=0.35]{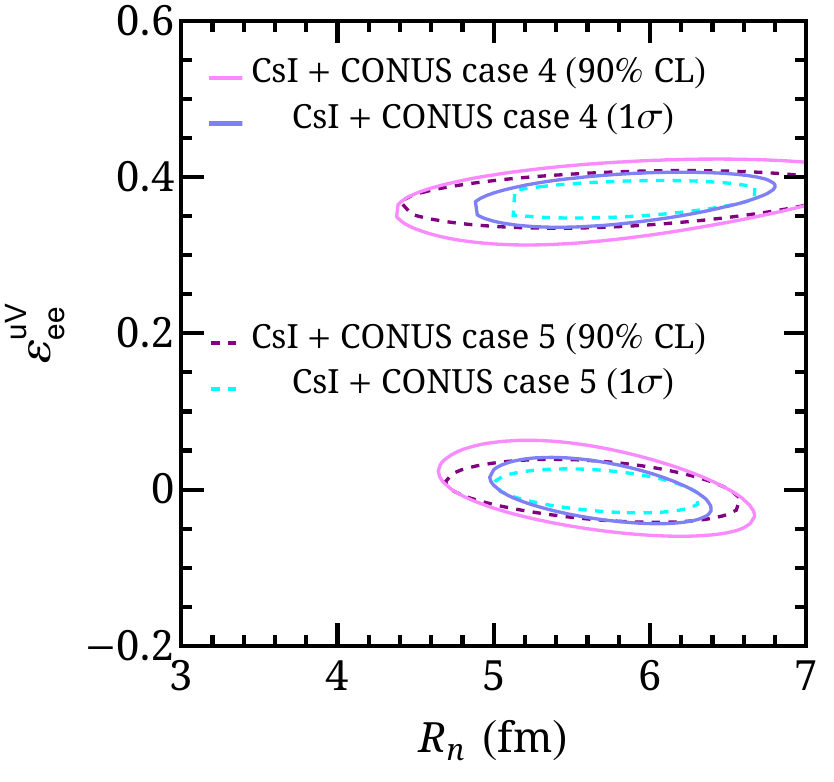}
     \includegraphics[scale=0.35]{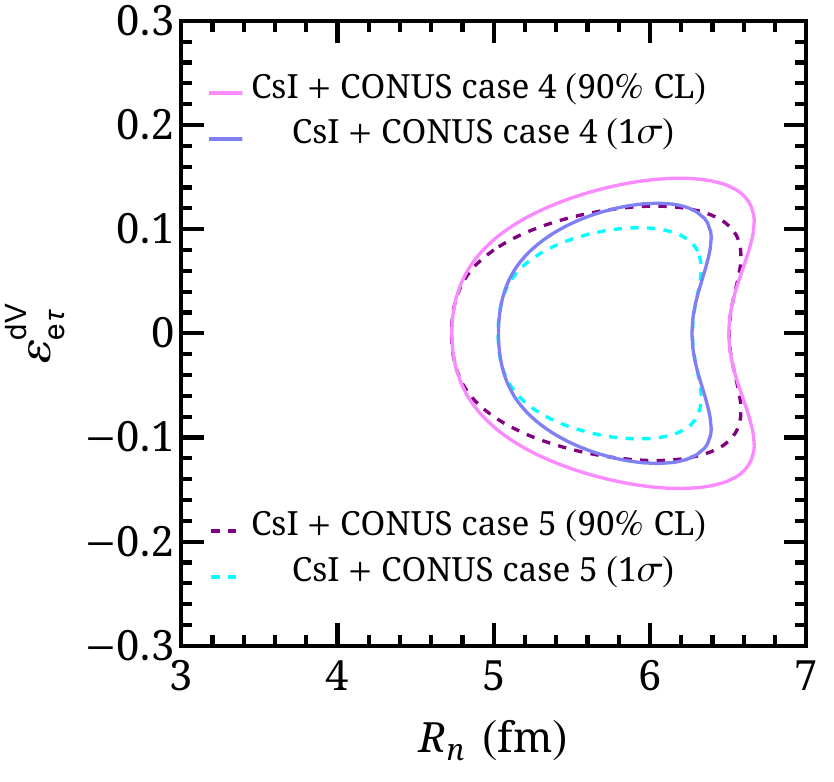}
    \caption{Top panels: Colored areas indicate the expected 1$\sigma$ (blue) and 90\% C.L. (pink) allowed regions from current CsI COHERENT data in the $(R_n,\varepsilon_{ee}^{dV})$, $(R_n,\varepsilon_{ee}^{uV})$, and $(R_n,\varepsilon_{e\tau}^{dV})$  planes. Dashed contours
    correspond to the combination with the case 1 CONUS-like reactor experiment setup at $1\sigma$ (blue lines) and 90\% C.L.  (black lines). Bottom panels: $1\sigma$ and 90\% C.L. allowed regions from the combination of CsI data and the scenarios 4 (solid) and 5 (dashed) presented in Tab.~\ref{tab:reac:char}. }
    \label{fig:eedV:rn:CsI}
\end{figure}

 We start by analyzing the currently available COHERENT CsI data allowing simultaneously for different values of the neutron rms radius, $R_n$, and the presence of neutrino non-standard interactions with matter.
The analysis follows the strategy presented in Sec.~\ref{sec:analysis}. Differently to the individual analysis in Secs.~\ref{sec:res:nucl} and \ref{sec:res:nsi:pidar}, here we evaluate the $\chi^2$ function in Eq.~(\ref{eq:chi:CsI}) varying at the same time the set of parameters $X = \{R_n,  ~\varepsilon_{\alpha\beta}^{qV}\}$.
\footnote{A similar analysis on the parameter space $(R_n,\varepsilon_{ee}^{dV})$ using the 2017 COHERENT CsI data set was performed in Ref.~\cite{canas2020interplay}.}
The results of our analysis are presented in Fig.~\ref{fig:eedV:rn:CsI}. Colored areas in the top panels show the allowed regions in the plane $(R_n,\varepsilon_{ee}^{dV})$ in the left panel, $(R_n,\varepsilon_{ee}^{uV})$ in the central panel, and $(R_n,\varepsilon_{e\tau}^{dV})$ in the right panel. 
The pink regions show the results at 90\% C.L. for two degrees of freedom. Notice that the presence of non-universal NSI allows for values of the neutron rms radius as low as $\approx$ 4.2 fm, while in the case of the flavor changing NSI, the rms radius is bounded from below at $\approx$ 4.7 fm.  This can be understood by considering that the presence of flavor-changing NSI always increases the expected number of events, while non-universal NSI interfere with the SM prediction and may increase or reduce the number of expected events. As a result, the value of $R_n$ is less constrained in the latter case.
For completeness, the colored blue regions in the figure show the results at the 1$\sigma$ level.

To investigate the interplay between NSI and nuclear parameters when using different neutrino sources, we combine the previous CsI results with the expected sensitivity of the reactor experiments described in the previous section to neutrino NSI. 
As a particular example, we combine COHERENT data with the CONUS-like reactor experiment studied above. 
Regions enclosed by dashed lines in the top panels of Fig.~\ref{fig:eedV:rn:CsI} show the results from the combination of CsI data and the CONUS-like detector with the experimental characteristics labeled as case 1 in Tab.~\ref{tab:reac:char}. Again, the pink color corresponds to 90\% C.L. and the blue color to 1$\sigma$ level. 
For the latter, one sees that, in all three panels, the combination results in a closed region around the SM solution. 
Additionally,  the bottom panels of the same figure explore the combination of the results assuming the more futuristic reactor scenarios indicated as case 4 (solid lines) and case 5 (dashed lines) in Tab.~\ref{tab:reac:char}. 
 Notice that, even at 90\% C.L., and in both cases, 4 and 5, we obtain closed regions in the parameter space in the three panels, making the determination of the neutron rms radius in the presence of NSI more robust from the combination of the two types of neutrino sources.

\begin{figure}[t]
    \centering
    \includegraphics[scale=0.35]{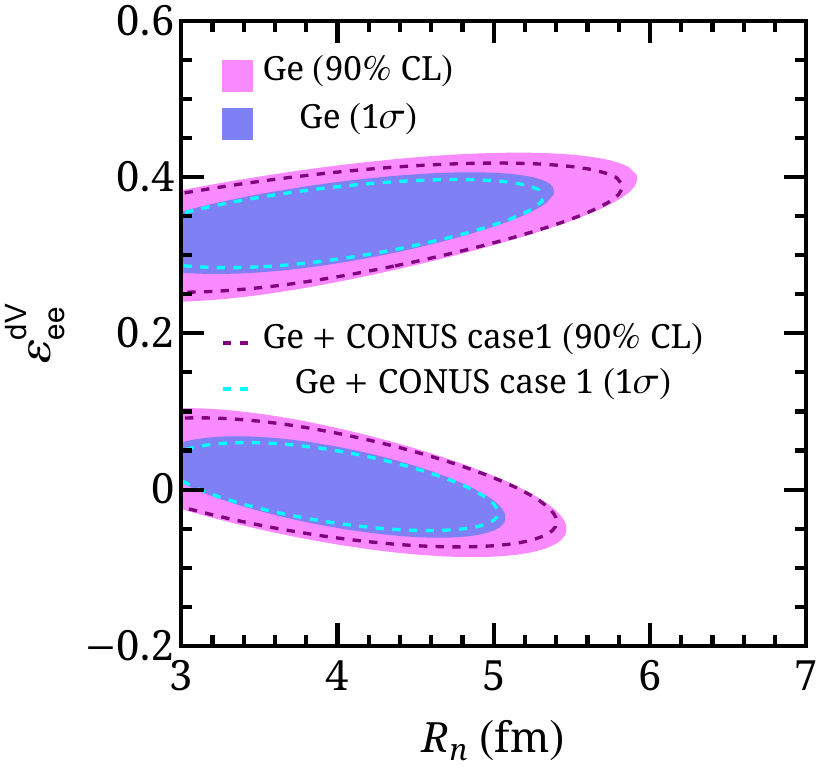}
    \includegraphics[scale=0.35]{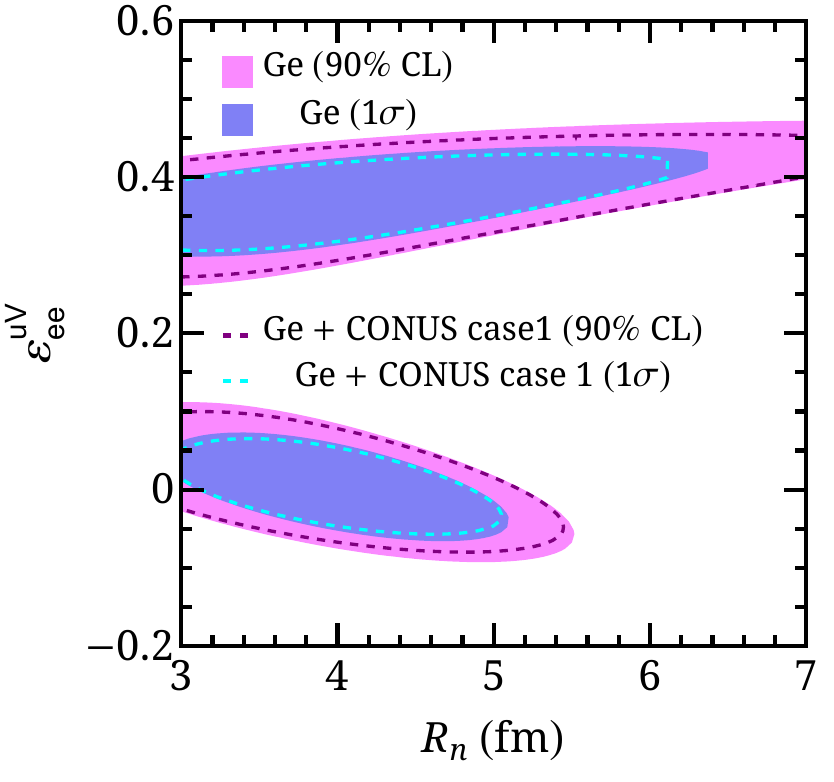}
    \includegraphics[scale=0.35]{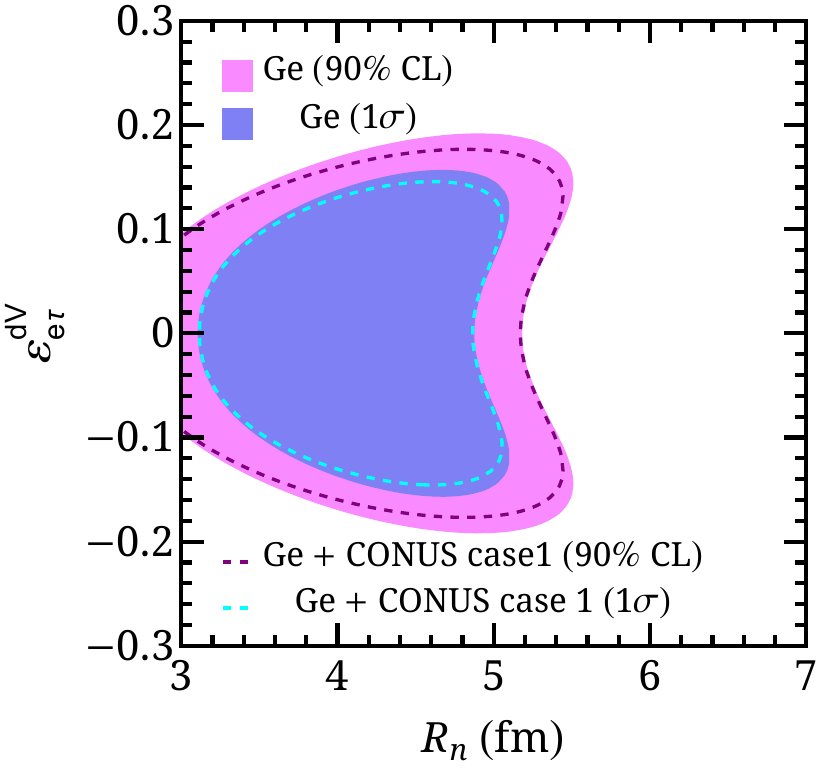}
    \includegraphics[scale=0.35]{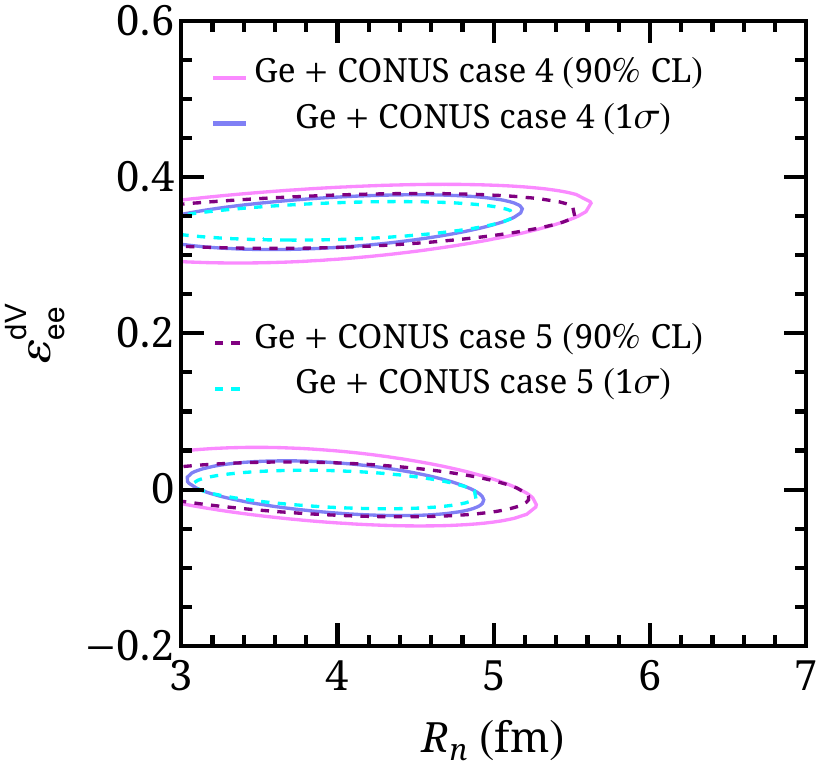}
    \includegraphics[scale=0.35]{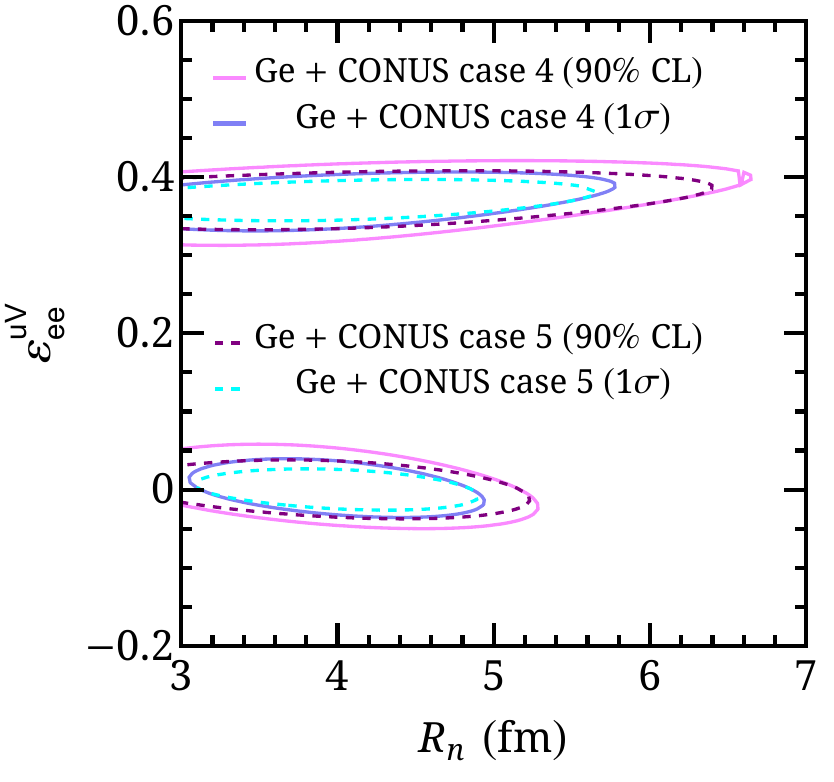}
    \includegraphics[scale=0.35]{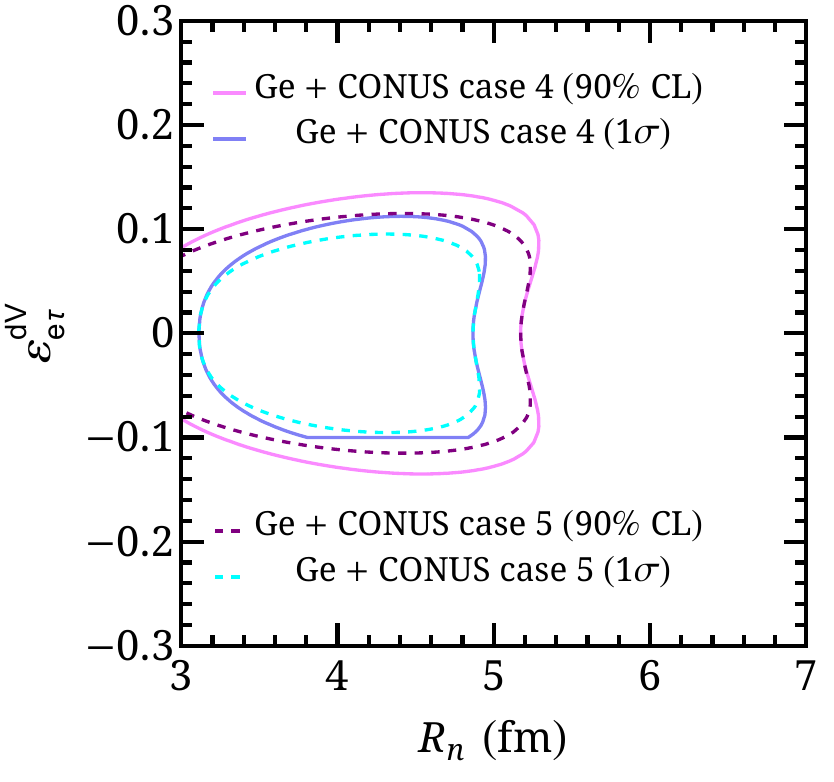}
    \caption{Top panels: Colored areas indicate the expected 1$\sigma$ (blue) and 90\% C.L. (pink) allowed regions from a future Ge detector at the SNS in the $(R_n,\varepsilon_{ee}^{dV})$, $(R_n,\varepsilon_{ee}^{uV})$, and $(R_n,\varepsilon_{e\tau}^{dV})$ planes. Dashed contours
  correspond to the combination with the case 1  CONUS-like reactor experiment setup at $1\sigma$ (blue lines) and 90\% C.L. (black lines). Bottom panels: $1\sigma$ and 90\% C.L. allowed regions from the combination of the Ge detector at the SNS and the cases 4 (solid) and 5 (dashed) presented in Tab.~\ref{tab:reac:char}. }
    \label{fig:eedV:rn:Ge:CONUS}
\end{figure}

 After trying to simultaneously constrain the neutron rms radius and neutrino NSI with current accelerator data, we now consider the expected sensitivity of the future $\pi$-DAR source detectors discussed in Sec.~\ref{sec:source} by minimizing the $\chi^2$ function defined in Eq.~(\ref{eq:chi:future}). 
The expected results assuming a $^{72}$Ge detector at the SNS are shown on the top panels of Fig.~\ref{fig:eedV:rn:Ge:CONUS} as blue  ($1\sigma$) and pink regions (90\% C.L.) in the planes  $(R_n,\varepsilon_{ee}^{dV})$ in the left panel, $(R_n,\varepsilon_{ee}^{uV})$ in the central panel, and $(R_n,\varepsilon_{e\tau}^{dV})$ in the right panel. 
For non-universal NSI, we can clearly see the presence of two separate regions, one around the SM model prediction, and another one preferring non-zero values for the NSI, and therefore allowing for larger values of the neutron rms radius. 
The results for flavour-changing NSI with $down$ quarks in the right panel show that the more $\varepsilon_{e\tau}^{dV}$ deviates from zero,  the larger is the allowed value for $R_n$.
The corresponding results for flavour-changing NSI with $up$ quarks are very similar and we do not show them here.

 As in the case of CsI, we now proceed to analyze the combined sensitivity of the future Ge detector at the SNS described in Tab.~\ref{tab:snschar} with our forecast for NSI searches at upcoming  CE$\nu$NS searches at reactor experiments.
Contours surrounded by dashed lines in the top panels of Fig.~\ref{fig:eedV:rn:Ge:CONUS} show the combination with the CONUS-like case 1 indicated in Tab. \ref{tab:reac:char}. Since for this case the reactor NSI constraints are not competitive,  the allowed region in each panel remains almost unchanged and the complementarity between the two experiments is not evident. The situation is different when we combine the SNS analysis with the most optimistic CONUS-like cases  4 and 5, listed in Tab. \ref{tab:reac:char}, as seen from the solid and dashed lines in the bottom panels of Fig.~\ref{fig:eedV:rn:Ge:CONUS}. 
 Here the improvement obtained with the combined analysis is clearly seen in the three panels, where the allowed region of $R_n$ is reduced and the sensitivity to the NSI couplings is dominated by reactor data.

\begin{figure}[t]
    \centering
    \includegraphics[scale=0.35]{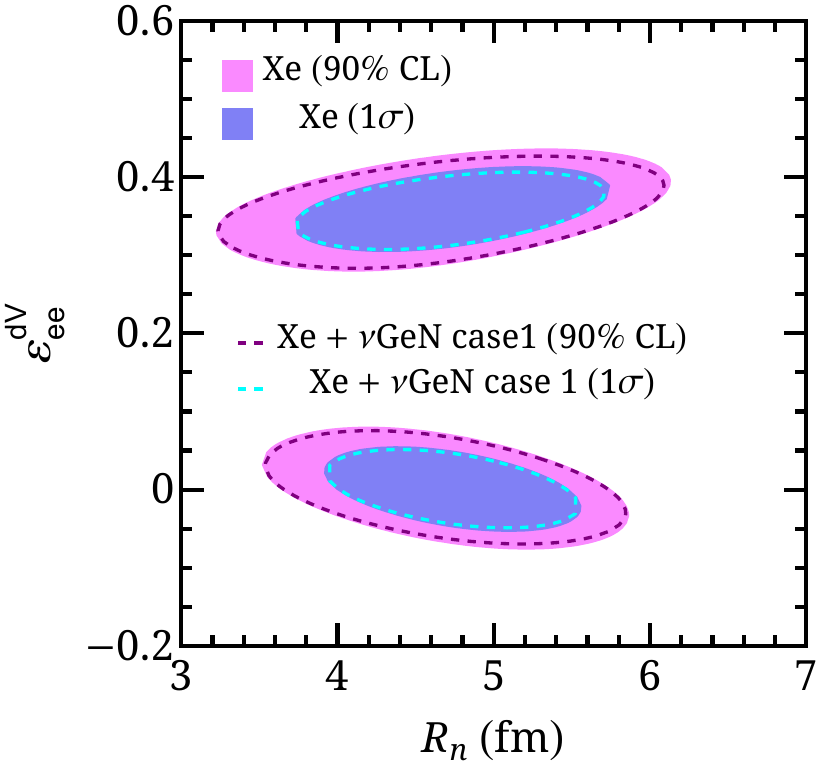}
    \includegraphics[scale=0.35]{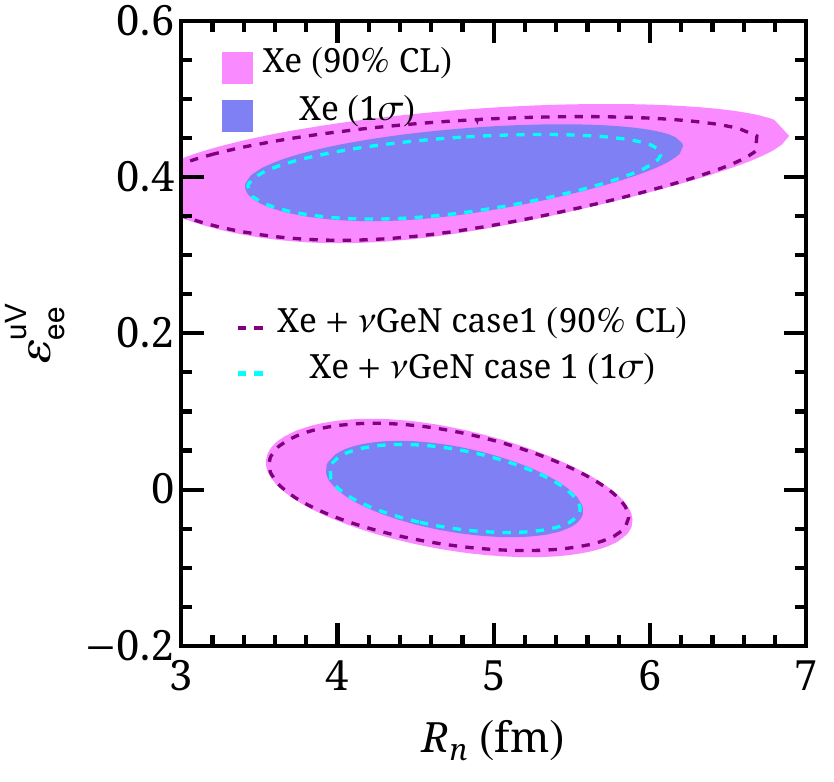}
    \includegraphics[scale=0.35]{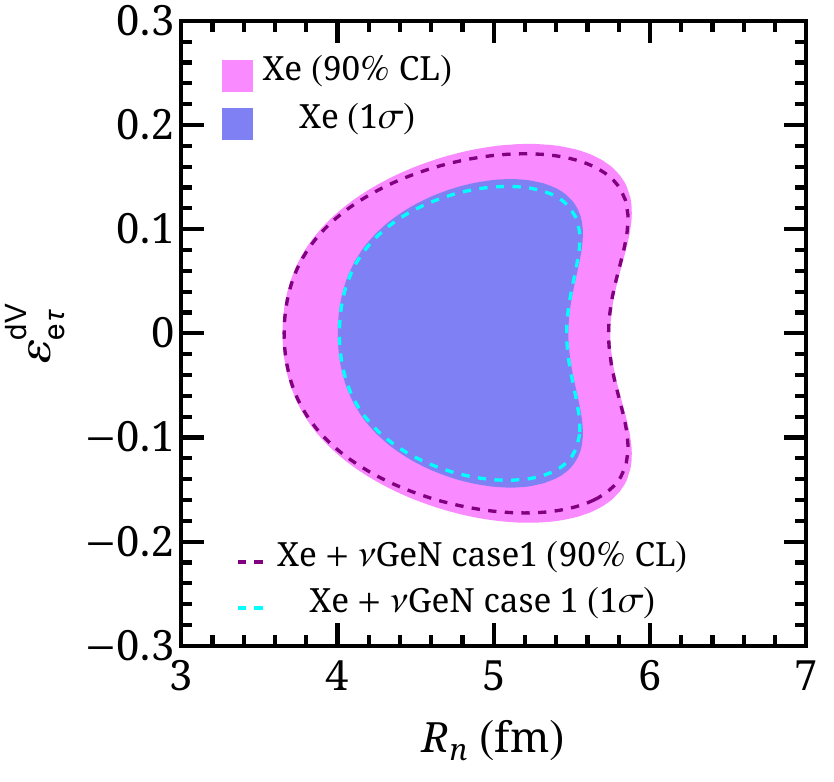}
    \includegraphics[scale=0.35]{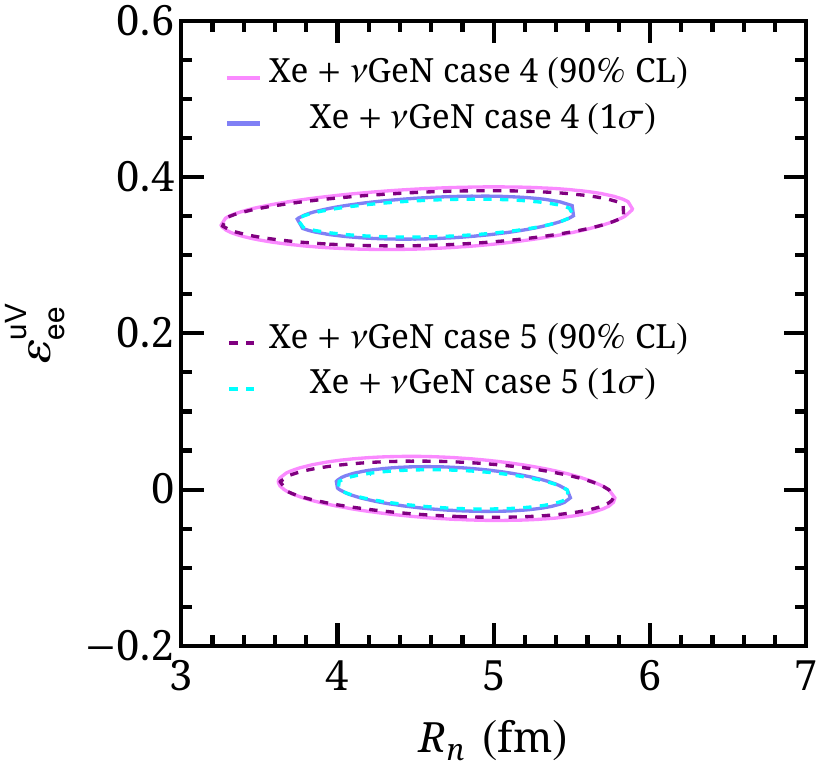}
    \includegraphics[scale=0.35]{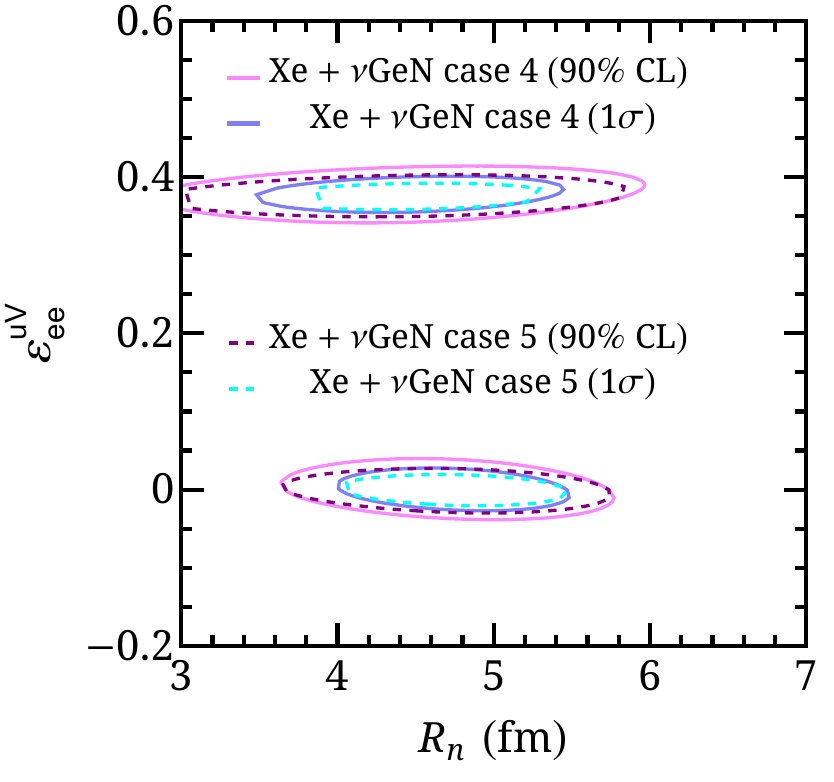}
    \includegraphics[scale=0.35]{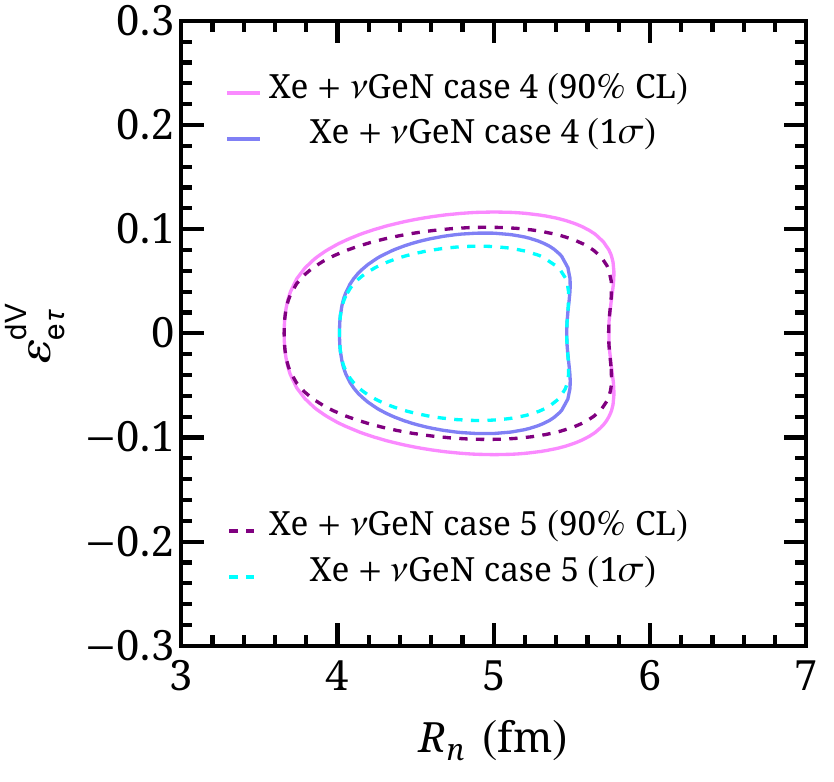}
    \caption{Top panels: Colored areas indicate the expected 1$\sigma$ (blue) and 90\% C.L. (pink) allowed regions from a Xe detector at the ESS in the $(R_n,\varepsilon_{ee}^{dV})$, $(R_n,\varepsilon_{ee}^{uV})$, and $(R_n,\varepsilon_{e\tau}^{dV})$ planes. Dashed contours correspond to the combination with the case 1 $\nu$GeN-like reactor experiment setup at $1\sigma$ (blue lines) and 90\% C.L. (black lines). Bottom panels: $1\sigma$ and 90\% C.L. allowed regions from the combination of the Xe detector at the ESS and the cases 4 (solid) and 5 (dashed) presented in Tab.~\ref{tab:reac:char}. }
    \label{fig:eedV:rn:Ge:nuGen}
\end{figure}

To conclude this section, we present the  joint analysis of the future Xe detector at the ESS described in Tab.~\ref{tab:snschar} and a future $\nu$GeN-like reactor detector. 
The results for the $\pi$-DAR Xe detector alone are displayed as coloured regions in the top panels of Fig.~\ref{fig:eedV:rn:Ge:nuGen}, with the same color code as in previous figures, for $\varepsilon_{ee}^{dV}$ (left), $\varepsilon_{ee}^{uV}$ (central), and $\varepsilon_{e\tau}^{dV}$ (right). Dashed lines in the same panels show the results of the combined analysis from a Xe detector with a  CE$\nu$NS reactor neutrino experiment $\nu$GeN-like detector, as described in Sec.~\ref{sec:source} and in case 1 of Tab.~\ref{tab:reac:char}. 
The combination with cases 4 and 5 is presented in the bottom panels of the figure with solid and dashed lines, respectively.
Note that the results presented in Figs.~\ref{fig:eedV:rn:Ge:CONUS} and \ref{fig:eedV:rn:Ge:nuGen} are not directly comparable since the chosen target materials at the SNS and ESS are different. However, we see some qualitative similarities between the two analyses, showing the complementarity between the CE$\nu$NS measurements from different neutrino sources.

\section{Conclusions}\label{sec:conclusions}

In this paper, we have studied the complementarity between CE$\nu$NS 
experiments that use different neutrino sources as well as different 
detection target materials to constrain nuclear uncertainties and new neutrino interactions with matter. 
Besides analyzing the sensitivity of current and future experimental setups to these physics cases individually, one of the main goals of this work consists of showing the synergies between
CE$\nu$NS experiments with $\pi$-DAR and reactor neutrino sources to simultaneously constrain the two scenarios.
This analysis is motivated by the fact that, at  CE$\nu$NS experiments with $\pi$-DAR sources, nuclear uncertainties parametrized in the neutron rms radius and neutrino NSI would produce a  similar impact on the observed neutrino signal. As a result, it would be impossible to disentangle the origin of a potential deviation with respect to the SM predicted signal and the sensitivity to new physics beyond the SM would be strongly affected by that.
 Indeed, we have shown that, when analyzing data from stopped pion sources alone,  the presence of nuclear uncertainties relevant for the CE$\nu$NS cross section might allow for rather large values of neutrino NSI couplings.
Fortunately, combined analyses of $\pi-$DAR  data with reactor CE$\nu$NS experiments, insensitive to nuclear uncertainties, can help break this degeneracy and thus provide more constraining results.

In our analysis, we have first estimated the sensitivity of stopped pion CE$\nu$NS experiments to nuclear physics and non-standard neutrino interactions separately, as well as the potential of CE$\nu$NS reactor neutrino experiments to constrain NSI.
As for $\pi$-DAR experiments,  we have 
considered the latest data from the CsI COHERENT experiment, the expected signal at the Ge COHERENT detector at the SNS, and a potential experiment using a Xe detector at the ESS.
In the reactor sector, we have proposed two experimental configurations based on optimized versions of the existing detectors CONUS and $\nu$GeN.
We have also estimated the sensitivity of $\pi$-DAR experiments to the two physics cases simultaneously, showing the presence of a degeneracy between  $R_n$ and neutrino NSI with matter, for both non-universal and flavor-changing interactions. This degeneracy is particularly relevant in the case of current CsI COHERENT data, although it also affects the sensitivity of the  Ge detector at the SNS and a potential Xe detector at the ESS. 
Even though the combination  with  the near-future sensitivity of a CONUS-like CE$\nu$NS reactor experiment 
can only slightly improve the sensitivity to the neutron rms radius, we have shown that a further improvement in the mass and threshold of the detector in a CONUS-like reactor experiment can have a significant impact on the determination of the neutron rms radius and the size of the NSI couplings. 
This illustrates the complementarity of 
CE$\nu$NS experiments using different neutrino sources to constrain the presence of new physics beyond the SM, even in the presence of nuclear uncertainties that could hinder this task. 
The choice of potential future experiments in this work is mainly based on the optimization of existing CE$\nu$NS experiments and could be regarded as a guide to enhance the experimental sensitivity to new physics in the presence of nuclear uncertainties.

\section*{Acknowledgements}

We thank Dimitris Papoulias and Omar Miranda for very fruitful discussions. This work was supported by the Spanish grants PID2020-113775GBI00  (MCIN/AEI/10.13039/501100011033), CIPROM/2021/054  and CIAPOS/2022/254 (Generalitat Valenciana).
R.R.R. has received support from the Coordenação de Aperfeiçoamento de Pessoal de Nível Superior - Brasil (CAPES) - Finance Code 001.

\bibliography{references.bib}

\begin{thebibliography}{10}

\bibitem{PhysRevD.9.1389}
D.~Z. Freedman, ``Coherent effects of a weak neutral current,'' {\em Phys. Rev.
  D}, vol.~9, pp.~1389--1392, Mar 1974.

\bibitem{COHERENT:2017ipa}
D.~Akimov {\em et~al.}, ``{Observation of Coherent Elastic Neutrino-Nucleus
  Scattering},'' {\em Science}, vol.~357, no.~6356, pp.~1123--1126, 2017.

\bibitem{COHERENT:2020iec}
D.~Akimov {\em et~al.}, ``{First Measurement of Coherent Elastic
  Neutrino-Nucleus Scattering on Argon},'' {\em Phys. Rev. Lett.}, vol.~126,
  no.~1, p.~012002, 2021.

\bibitem{COHERENT:2021xmm}
D.~Akimov {\em et~al.}, ``{Measurement of the Coherent Elastic Neutrino-Nucleus
  Scattering Cross Section on CsI by COHERENT},'' {\em Phys. Rev. Lett.},
  vol.~129, no.~8, p.~081801, 2022.

\bibitem{AtzoriCorona:2023ktl}
M.~Atzori~Corona, M.~Cadeddu, N.~Cargioli, F.~Dordei, C.~Giunti, and G.~Masia,
  ``{Nuclear neutron radius and weak mixing angle measurements from latest
  COHERENT CsI and atomic parity violation Cs data},'' 3 2023.

\bibitem{Cadeddu:2021ijh}
M.~Cadeddu, N.~Cargioli, F.~Dordei, C.~Giunti, Y.~F. Li, E.~Picciau, C.~A.
  Ternes, and Y.~Y. Zhang, ``{New insights into nuclear physics and weak mixing
  angle using electroweak probes},'' {\em Phys. Rev. C}, vol.~104, no.~6,
  p.~065502, 2021.

\bibitem{Cadeddu:2017etk}
M.~Cadeddu, C.~Giunti, Y.~F. Li, and Y.~Y. Zhang, ``{Average CsI neutron
  density distribution from COHERENT data},'' {\em Phys. Rev. Lett.}, vol.~120,
  no.~7, p.~072501, 2018.

\bibitem{Papoulias:2019lfi}
D.~K. Papoulias, T.~S. Kosmas, R.~Sahu, V.~K.~B. Kota, and M.~Hota,
  ``{Constraining nuclear physics parameters with current and future COHERENT
  data},'' {\em Phys. Lett. B}, vol.~800, p.~135133, 2020.

\bibitem{Ohlsson:2012kf}
T.~Ohlsson, ``{Status of non-standard neutrino interactions},'' {\em Rept.
  Prog. Phys.}, vol.~76, p.~044201, 2013.

\bibitem{Denton:2020hop}
P.~B. Denton and J.~Gehrlein, ``{A Statistical Analysis of the COHERENT Data
  and Applications to New Physics},'' {\em JHEP}, vol.~04, p.~266, 2021.

\bibitem{Miranda:2020tif}
O.~G. Miranda, D.~K. Papoulias, G.~Sanchez~Garcia, O.~Sanders, M.~T\'ortola,
  and J.~W.~F. Valle, ``{Implications of the first detection of coherent
  elastic neutrino-nucleus scattering (CEvNS) with Liquid Argon},'' {\em JHEP},
  vol.~05, p.~130, 2020.
\newblock [Erratum: JHEP 01, 067 (2021)].

\bibitem{Giunti:2019xpr}
C.~Giunti, ``{General COHERENT constraints on neutrino nonstandard
  interactions},'' {\em Phys. Rev. D}, vol.~101, no.~3, p.~035039, 2020.

\bibitem{Miranda:2019skf}
O.~G. Miranda, G.~Sanchez~Garcia, and O.~Sanders, ``{Coherent elastic
  neutrino-nucleus scattering as a precision test for the Standard Model and
  beyond: the COHERENT proposal case},'' {\em Adv. High Energy Phys.},
  vol.~2019, p.~3902819, 2019.
\newblock [Erratum: Adv.High Energy Phys. 2022, 9874517 (2022)].

\bibitem{Miranda:2019wdy}
O.~G. Miranda, D.~K. Papoulias, M.~T\'ortola, and J.~W.~F. Valle, ``{Probing
  neutrino transition magnetic moments with coherent elastic neutrino-nucleus
  scattering},'' {\em JHEP}, vol.~07, p.~103, 2019.

\bibitem{Kosmas:2015sqa}
T.~S. Kosmas, O.~G. Miranda, D.~K. Papoulias, M.~Tortola, and J.~W.~F. Valle,
  ``{Probing neutrino magnetic moments at the Spallation Neutron Source
  facility},'' {\em Phys. Rev. D}, vol.~92, no.~1, p.~013011, 2015.

\bibitem{Cadeddu:2018dux}
M.~Cadeddu, C.~Giunti, K.~A. Kouzakov, Y.-F. Li, Y.-Y. Zhang, and A.~I.
  Studenikin, ``{Neutrino Charge Radii From Coherent Elastic Neutrino-nucleus
  Scattering},'' {\em Phys. Rev. D}, vol.~98, no.~11, p.~113010, 2018.
\newblock [Erratum: Phys.Rev.D 101, 059902 (2020)].

\bibitem{Parada:2019gvy}
A.~Parada, ``{Constraints on neutrino electric millicharge from experiments of
  elastic neutrino-electron interaction and future experimental proposals
  involving coherent elastic neutrino-nucleus scattering},'' {\em Adv. High
  Energy Phys.}, vol.~2020, p.~5908904, 2020.

\bibitem{Lindner:2016wff}
M.~Lindner, W.~Rodejohann, and X.-J. Xu, ``{Coherent Neutrino-Nucleus
  Scattering and new Neutrino Interactions},'' {\em JHEP}, vol.~03, p.~097,
  2017.

\bibitem{AristizabalSierra:2018eqm}
D.~Aristizabal~Sierra, V.~De~Romeri, and N.~Rojas, ``{COHERENT analysis of
  neutrino generalized interactions},'' {\em Phys. Rev. D}, vol.~98, p.~075018,
  2018.

\bibitem{Flores:2021kzl}
L.~J. Flores, N.~Nath, and E.~Peinado, ``{CE\ensuremath{\nu}NS as a probe of
  flavored generalized neutrino interactions},'' {\em Phys. Rev. D}, vol.~105,
  no.~5, p.~055010, 2022.

\bibitem{AristizabalSierra:2019ufd}
D.~Aristizabal~Sierra, V.~De~Romeri, and N.~Rojas, ``{CP violating effects in
  coherent elastic neutrino-nucleus scattering processes},'' {\em JHEP},
  vol.~09, p.~069, 2019.

\bibitem{Abdullah:2018ykz}
M.~Abdullah, J.~B. Dent, B.~Dutta, G.~L. Kane, S.~Liao, and L.~E. Strigari,
  ``{Coherent elastic neutrino nucleus scattering as a probe of a Z' through
  kinetic and mass mixing effects},'' {\em Phys. Rev. D}, vol.~98, no.~1,
  p.~015005, 2018.

\bibitem{Cadeddu:2020nbr}
M.~Cadeddu, N.~Cargioli, F.~Dordei, C.~Giunti, Y.~F. Li, E.~Picciau, and Y.~Y.
  Zhang, ``{Constraints on light vector mediators through coherent elastic
  neutrino nucleus scattering data from COHERENT},'' {\em JHEP}, vol.~01,
  p.~116, 2021.

\bibitem{Flores:2020lji}
L.~J. Flores, N.~Nath, and E.~Peinado, ``{Non-standard neutrino interactions in
  U(1)' model after COHERENT data},'' {\em JHEP}, vol.~06, p.~045, 2020.

\bibitem{Kosmas:2017zbh}
T.~S. Kosmas, D.~K. Papoulias, M.~Tortola, and J.~W.~F. Valle, ``{Probing light
  sterile neutrino signatures at reactor and Spallation Neutron Source neutrino
  experiments},'' {\em Phys. Rev. D}, vol.~96, no.~6, p.~063013, 2017.

\bibitem{Blanco:2019vyp}
C.~Blanco, D.~Hooper, and P.~Machado, ``{Constraining Sterile Neutrino
  Interpretations of the LSND and MiniBooNE Anomalies with Coherent Neutrino
  Scattering Experiments},'' {\em Phys. Rev. D}, vol.~101, no.~7, p.~075051,
  2020.

\bibitem{Candela:2023rvt}
P.~M. Candela, V.~De~Romeri, and D.~K. Papoulias, ``{COHERENT production of a
  Dark Fermion},'' 5 2023.

\bibitem{DeRomeri:2022twg}
V.~De~Romeri, O.~G. Miranda, D.~K. Papoulias, G.~Sanchez~Garcia, M.~T\'ortola,
  and J.~W.~F. Valle, ``{Physics implications of a combined analysis of
  COHERENT CsI and LAr data},'' {\em JHEP}, vol.~04, p.~035, 2023.

\bibitem{COHERENT:2021yvp}
D.~Akimov {\em et~al.}, ``{Simulating the neutrino flux from the Spallation
  Neutron Source for the COHERENT experiment},'' {\em Phys. Rev. D}, vol.~106,
  no.~3, p.~032003, 2022.

\bibitem{CCM:2021leg}
A.~A. Aguilar-Arevalo {\em et~al.}, ``{First dark matter search results from
  Coherent CAPTAIN-Mills},'' {\em Phys. Rev. D}, vol.~106, no.~1, p.~012001,
  2022.

\bibitem{Baxter:2019mcx}
D.~Baxter {\em et~al.}, ``{Coherent Elastic Neutrino-Nucleus Scattering at the
  European Spallation Source},'' {\em JHEP}, vol.~02, p.~123, 2020.

\bibitem{Akimov:2022oyb}
D.~Akimov {\em et~al.}, ``{The COHERENT Experimental Program},'' in {\em
  {Snowmass 2021}}, 4 2022.

\bibitem{Galindo-Uribarri:2020huw}
A.~Galindo-Uribarri, O.~G. Miranda, and G.~S. Garcia, ``{Novel approach for the
  study of coherent elastic neutrino-nucleus scattering},'' {\em Phys. Rev. D},
  vol.~105, no.~3, p.~033001, 2022.

\bibitem{chatterjee2023constraining}
S.~S. Chatterjee, S.~Lavignac, O.~Miranda, and G.~S. Garcia, ``Constraining
  nonstandard interactions with coherent elastic neutrino-nucleus scattering at
  the european spallation source,'' {\em Physical Review D}, vol.~107, no.~5,
  p.~055019, 2023.

\bibitem{vonRaesfeld:2021gxl}
C.~von Raesfeld and P.~Huber, ``{Use of CEvNS to monitor spent nuclear fuel},''
  {\em Phys. Rev. D}, vol.~105, no.~5, p.~056002, 2022.

\bibitem{Colaresi:2022obx}
J.~Colaresi, J.~I. Collar, T.~W. Hossbach, C.~M. Lewis, and K.~M. Yocum,
  ``{Measurement of Coherent Elastic Neutrino-Nucleus Scattering from Reactor
  Antineutrinos},'' {\em Phys. Rev. Lett.}, vol.~129, no.~21, p.~211802, 2022.

\bibitem{CONUS:2020skt}
H.~Bonet {\em et~al.}, ``{Constraints on elastic neutrino nucleus scattering in
  the fully coherent regime from the CONUS experiment},'' {\em Phys. Rev.
  Lett.}, vol.~126, no.~4, p.~041804, 2021.

\bibitem{nGeN:2022uje}
I.~Alekseev {\em et~al.}, ``{First results of the \ensuremath{\nu}GeN
  experiment on coherent elastic neutrino-nucleus scattering},'' {\em Phys.
  Rev. D}, vol.~106, no.~5, p.~L051101, 2022.

\bibitem{CONNIE:2021ggh}
A.~Aguilar-Arevalo {\em et~al.}, ``{Search for coherent elastic
  neutrino-nucleus scattering at a nuclear reactor with CONNIE 2019 data},''
  {\em JHEP}, vol.~05, p.~017, 2022.

\bibitem{Akimov:2022xvr}
D.~Y. Akimov {\em et~al.}, ``{The RED-100 experiment},'' {\em JINST}, vol.~17,
  no.~11, p.~T11011, 2022.

\bibitem{canas2020interplay}
B.~Canas, E.~Garces, O.~Miranda, A.~Parada, and G.~S. Garcia, ``Interplay
  between nonstandard and nuclear constraints in coherent elastic
  neutrino-nucleus scattering experiments,'' {\em Physical Review D}, vol.~101,
  no.~3, p.~035012, 2020.

\bibitem{Tomalak:2020zfh}
O.~Tomalak, P.~Machado, V.~Pandey, and R.~Plestid, ``{Flavor-dependent
  radiative corrections in coherent elastic neutrino-nucleus scattering},''
  {\em JHEP}, vol.~02, p.~097, 2021.

\bibitem{Piekarewicz:2016vbn}
J.~Piekarewicz, A.~R. Linero, P.~Giuliani, and E.~Chicken, ``{Power of two:
  Assessing the impact of a second measurement of the weak-charge form factor
  of $^{208}$Pb},'' {\em Phys. Rev. C}, vol.~94, no.~3, p.~034316, 2016.

\bibitem{PhysRev.104.1466}
R.~H. Helm, ``Inelastic and elastic scattering of 187-mev electrons from
  selected even-even nuclei,'' {\em Phys. Rev.}, vol.~104, pp.~1466--1475, Dec
  1956.

\bibitem{PhysRevC.60.014903}
S.~R. Klein and J.~Nystrand, ``Exclusive vector meson production in
  relativistic heavy ion collisions,'' {\em Phys. Rev. C}, vol.~60, p.~014903,
  Jun 1999.

\bibitem{Sierra:2023pnf}
D.~A. Sierra, ``{Extraction of neutron density distributions from
  high-statistics coherent elastic neutrino-nucleus scattering data},'' 1 2023.

\bibitem{Angeli:2013epw}
I.~Angeli and K.~P. Marinova, ``{Table of experimental nuclear ground state
  charge radii: An update},'' {\em Atom. Data Nucl. Data Tabl.}, vol.~99,
  no.~1, pp.~69--95, 2013.

\bibitem{PhysRevD.17.2369}
L.~Wolfenstein, ``Neutrino oscillations in matter,'' {\em Phys. Rev. D},
  vol.~17, pp.~2369--2374, May 1978.

\bibitem{Barranco:2005yy}
J.~Barranco, O.~G. Miranda, and T.~I. Rashba, ``{Probing new physics with
  coherent neutrino scattering off nuclei},'' {\em JHEP}, vol.~12, p.~021,
  2005.

\bibitem{Farzan:2017xzy}
Y.~Farzan and M.~Tortola, ``{Neutrino oscillations and Non-Standard
  Interactions},'' {\em Front. in Phys.}, vol.~6, p.~10, 2018.

\bibitem{Picciau:2022xzi}
E.~Picciau, {\em {Low-energy signatures in DarkSide-50 experiment and neutrino
  scattering processes}}.
\newblock PhD thesis, Cagliari U., 2022.

\bibitem{fnalNeutrino2020}
K.~Mann, ``Measuring coherent neutrino scattering with ge.''
  \url{https://indico.fnal.gov/event/19348/contributions/186707/}, 2020.
\newblock [Accessed 23-Jun-2023].

\bibitem{Abele:2022iml}
H.~Abele {\em et~al.}, ``{Particle Physics at the European Spallation
  Source},'' 11 2022.

\bibitem{Kopeikin:2012zz}
V.~I. Kopeikin, ``{Flux and spectrum of reactor antineutrinos},'' {\em Phys.
  Atom. Nucl.}, vol.~75, pp.~143--152, 2012.

\bibitem{PhysRevC.84.024617}
P.~Huber, ``Determination of antineutrino spectra from nuclear reactors,'' {\em
  Phys. Rev. C}, vol.~84, p.~024617, Aug 2011.

\bibitem{Mueller:2011nm}
T.~A. Mueller {\em et~al.}, ``{Improved Predictions of Reactor Antineutrino
  Spectra},'' {\em Phys. Rev. C}, vol.~83, p.~054615, 2011.

\bibitem{Kopeikin:2004cn}
V.~Kopeikin, L.~Mikaelyan, and V.~Sinev, ``{Reactor as a source of
  antineutrinos: Thermal fission energy},'' {\em Phys. Atom. Nucl.}, vol.~67,
  pp.~1892--1899, 2004.

\bibitem{Buck:2020opf}
C.~Buck {\em et~al.}, ``{A novel experiment for coherent elastic neutrino
  nucleus scattering: CONUS},'' {\em J. Phys. Conf. Ser.}, vol.~1342, no.~1,
  p.~012094, 2020.

\bibitem{magcevns:edgar}
E.~Sanchez, ``The conus+ experiment.'' \url{https://indi.to/TRz9Z}, 2023.
\newblock [Accessed 13-Oct-2023].

\bibitem{CONUS:2021dwh}
H.~Bonet {\em et~al.}, ``{Novel constraints on neutrino physics beyond the
  standard model from the CONUS experiment},'' {\em JHEP}, vol.~05, p.~085,
  2022.

\bibitem{Pershey:M7s}
D.~Pershey, ``New results from the coherent csi[na] detector. (talk at
  magnificent cevns).'' \url{
  "https://indico.cern.ch/event/943069/contributions/4066386/"}, 2020.

\bibitem{Horowitz:2012tj}
C.~J. Horowitz {\em et~al.}, ``{Weak charge form factor and radius of 208Pb
  through parity violation in electron scattering},'' {\em Phys. Rev. C},
  vol.~85, p.~032501, 2012.

\bibitem{CREX:2022kgg}
D.~Adhikari {\em et~al.}, ``{Precision Determination of the Neutral Weak Form
  Factor of Ca48},'' {\em Phys. Rev. Lett.}, vol.~129, no.~4, p.~042501, 2022.

\end{thebibliography}
\bibliographystyle{ieeetr}

\end{document}